%% file: seal-arxiv.tex
\documentclass{sig-alternate}
\usepackage{mathptmx} 
\usepackage{fancyhdr}
\usepackage[normalem]{ulem}
\usepackage[hyphens]{url}
\usepackage[sort,nocompress]{cite}
\usepackage[final]{microtype}

\usepackage{subfig}
\usepackage{comment}
\usepackage{multirow}
\usepackage{bm}
\usepackage{makecell}
\usepackage[bookmarks=true,breaklinks=true,letterpaper=true,colorlinks,linkcolor=red,citecolor=blue,urlcolor=black]{hyperref}

\newcommand{\drafts}[1]{\textcolor{black}{#1}}

\newcommand{\draftsnew}[1]{\textcolor{black}{#1}}

\newcommand{\postmicro}[1]{\textcolor{black}{#1}}

\newcommand{\isca}[1]{\textcolor{black}{#1}}

\newcommand{\iscare}[1]{\textcolor{black}{#1}}

\title{SEALing Neural Network Models in Secure Deep Learning Accelerators~\vspace{-1cm}}

\author{
	{   Pengfei Zuo{\small \textsuperscript{*}}{\small \textsuperscript{\textdagger}},
		Yu Hua{\small \textsuperscript{*}},
		Ling Liang{\small \textsuperscript{\textdagger}},
		Xinfeng Xie{\small \textsuperscript{\textdagger}},
		Xing Hu{\small \textsuperscript{\textdagger}},
		Yuan Xie{\small \textsuperscript{\textdagger}}
	}\\
	{\small \textsuperscript{*}}Huazhong University of Science and Technology \\
	{\small \textsuperscript{\textdagger}} Scalable Energy-efficient Architecture Lab (SEAL), University of California, Santa Barbara\\
}

\begin{document}
\maketitle
\pagestyle{plain}


\begin{abstract}

Deep learning (DL) accelerators are increasingly deployed on edge devices to support fast local inferences. However, they suffer from a new security problem, i.e., being vulnerable to physical access based attacks. An adversary can easily obtain the entire neural network (NN) model by physically snooping the GDDR (graphics double data rate) memory bus that connects the accelerator chip with DRAM memory. Therefore, memory encryption becomes important for DL accelerators on edge devices to improve the security of NN models. Nevertheless, we observe that traditional memory encryption solutions that have been efficiently used in CPU systems cause significant performance degradation when directly used in DL accelerators. The main reason comes from the big bandwidth gap between the GDDR memory bus and the encryption engine. To address this problem, our paper proposes SEAL, a Secure and Efficient Accelerator scheme for deep Learning. SEAL enhances the performance of the encrypted DL accelerator from two aspects, i.e., improving the data access bandwidth and the efficiency of memory encryption. Specifically, to improve the data access bandwidth, SEAL leverages a criticality-aware smart encryption scheme which identifies partial data that have no impact on the security of NN models and allows them to bypass the encryption engine, thus reducing the amount of data to be encrypted. To improve the efficiency of memory encryption, SEAL leverages a colocation mode encryption scheme to eliminate memory accesses from counters used for encryption by co-locating data and their counters. Our experimental results demonstrate that, compared with traditional memory encryption solutions, SEAL achieves $1.4\times-1.6 \times$ IPC improvement and reduces the inference latency by $39\%-60\%$. Compared with a baseline accelerator without memory encryption, SEAL compromises only $5\%-7\%$ IPC for significant security improvement.

\end{abstract}

\input{introduction.tex}

\input{background.tex}
\input{design.tex}
\input{evaluation.tex}
\input{relatedwork.tex}

\section{Conclusion}

Our paper proposes SEAL to enhance the security of NN models in DL accelerators on edge devices. To reduce performance overheads from memory encryption, SEAL leverages a criticality-aware smart encryption (SE) scheme and a colocation mode encryption (ColoE) scheme. The SE scheme is used to improve the data access bandwidth of DL accelerators by identifying partial data that have no impact on the security of NN models and allowing them to bypass the encryption engine without affecting the security of the NN model. The ColoE scheme is used to improve the efficiency of memory encryption by co-locating data and their counters to reduce the memory accesses from counters. Our experimental results show that, compared with traditional memory encryption solutions, SEAL achieves $1.4\times-1.6 \times$ IPC improvement. Compared with a baseline accelerator without using memory encryption, SEAL \postmicro{improves the} security with a slight overhead ( $5\%-7\%$ IPC).


\bibliographystyle{IEEEtranS}
\bibliography{references}

\end{document}

%% file: introduction.tex
\vspace{-8px}
\section{Introduction}

Machine learning techniques, especially deep learning (DL), have made significant progress in the past few years, whose performances have surpassed those of humans in some application domains, such as image classification~\cite{krizhevsky2012imagenet,simonyan2014very,he2016deep}, speech recognition~\cite{8049322,graves2013speech,cho2014learning}, and games~\cite{silver2016mastering}. With the increase of computing performance and storage capacity of edge devices, DL systems are increasingly expanded and used from cloud to edge devices~\cite{wu2019machine,edgeTPU2018}, such as self-driving cars~\cite{huval2015empirical} and Internet-of-things devices~\cite{li2018learning}. \draftsnew{By employing DL accelerators, e.g., GPU and NPU, edge devices are able to carry out real-time local inferences based on current environments without a connection with a remote control center with high latency.} For example, over 99\% smartphones are equipped with a GPU by 2019~\cite{scientiamobile2019,gpurank2020}. The self-driving computer within Tesla cars~\cite{tesla2019} and Google edge TPU~\cite{edgetpu2019} also include a GPU.

In DL accelerators, neural network (NN) models are confidential information that needs to be protected. Because NN models represent the Intellectual Property (IP) of model owners, which should be confidentially protected to preserve their competitive advantages. More importantly, the knowledge of NN models can facilitate an adversary to carry out more powerful adversarial attacks~\cite{szegedy2014intriguing,goodfellow2014generative}. In adversarial attacks, an adversary is able to intentionally affect the outcome of the DL inference by modifying the input data with a slight perturbation that is imperceptible to humans. For example, by performing adversarial attacks, an adversary is able to manipulate self-driving cars~\cite{eykholt2018robust} and trick the speaker recognition system in smartphones~\cite{carlini2016hidden}. In general, if the adversary does not know the NN model, the success rate of the adversarial attack is low. With the knowledge of the NN model, the success rate is significantly improved~\cite{liu2017delving,oh2018towards}.

However, DL accelerators deployed on edge devices suffer from a new security issue compared with those deployed on the cloud. The reason is that DL accelerators on edge devices are easier to be physically accessed, thus being vulnerable to physical access based attacks. \iscare{The accelerator chip and DRAM themselves are usually well packaged and hence secure to physical access, but the memory bus collecting accelerator and DRAM is not secure, due to being vulnerable to bus snooping attacks~\cite{Yan2006ICP,henson2014memory,hua2018reverse,hu2019neural}.} Since the DL accelerator has to access the NN model stored in the DRAM memory through the GDDR memory bus during the inference, an adversary can easily obtain the entire NN model by inserting a bus snooper on the GDDR bus to intercept the data communicated between the DL accelerator chip and the DRAM memory.
Therefore, memory encryption for encrypting the data transmitted between the DL accelerator chip and the DRAM memory is important.

There are two existing memory encryption models in secure CPU systems including direct encryption and counter mode encryption. Direct encryption encrypts all memory lines by using the same global key. It has a low security level since the same data are always encrypted to the same ciphertext, leaving the direct encryption vulnerable to dictionary and retry attacks~\cite{awad2016silent}. Counter mode encryption~\cite{lipmaa2000ctr} encrypts a memory line by using a globe key in conjunction with its line address and a per-line counter. Thus the same plaintexts are encrypted to different ciphertexts, achieving a high security level. Counter mode encryption needs to maintain a counter cache on the CPU chip. When accessing a memory line, if its corresponding counter is found in the counter cache, its decryption latency is hidden in the memory read latency to improve the system performance. \draftsnew{The reason is that counter mode encryption generates a one-time pad (OTP) using the counter in parallel with the memory read and decrypts the memory line by XORing the OTP with the data.} Due to the benefit of hiding decryption latency, counter mode encryption only incurs about $5\%$ performance overhead in CPU systems~\cite{Yan2006ICP}.

However, we observe that employing these memory encryption techniques in DL accelerators significantly decreases their performance. The IPC (instruction per cycle) of the DL accelerator is reduced by over $50\%$ after using memory encryption, as evaluated in Section~\ref{sec:background-straightforward-solution}. Such a significant performance decrease is unacceptable for the latency-sensitive DL accelerators on edge devices that must carry out real-time inferences based on current environments, e.g., self-driving cars. The main reason comes from the big bandwidth gap between the GDDR memory bus and the encryption engine. For DL accelerators, e.g., GPUs, their performance is highly bandwidth-bounded and hence the GDDR memory is designed for GPUs to achieve high memory access bandwidth. The bandwidth of the GDDR memory bus is generally higher than 160GB/s~\cite{Nvidia2018Turing,Nvidia2014GTX980,Nvidia2017GTX,Nvidia2012GTX480}. However, the state-of-the-art encryption engine with hardware implementation achieves only \iscare{about 8GB/s of bandwidth on average}~\cite{mathew201053gbps,Ensilica2020,sayilar2014cryptoraptor,liu2011parallel,morioka200410}. Even though we deploy one encryption engine in every memory controller, the big bandwidth gap remains. As a result, the high bandwidth of the GDDR memory bus is under-utilized and the encryption engine becomes the bandwidth bottleneck in secure DL accelerators. \postmicro{Moreover, since the data access bandwidth is the performance bottleneck, counter mode encryption causing extra memory accesses from counters exacerbates the performance on DL accelerators, which even delivers worse performance than direct encryption.}

To address these problems, our paper proposes SEAL\footnote{SEAL means we seal NN models in secure DL accelerators and thus no one can snoop them.}, a Secure and Efficient Accelerator scheme for deep Learning to enhance the security of DL accelerators on edge devices while delivering a high performance. SEAL reduces the performance overhead of encryption by using a criticality-aware smart encryption (SE) scheme and a colocation mode encryption (ColoE) scheme. Specifically, \drafts{to improve the data access bandwidth of DL accelerators, SEAL leverages the SE scheme to identify partial data having no impact on the security of NN models and allow them to bypass the encryption engine, lowering the amount of data to be encrypted without affecting security.} To improve the efficiency of memory encryption, SEAL leverages the ColoE scheme that co-locates the storage of each data and its counter. ColoE has the same security level as the traditional counter mode encryption while achieving higher performance in DL accelerators due to removing extra memory accesses from counters. In summary, this paper makes the following contributions.

  \textbf{\emph{$\bullet$ Observations and Insights on Securing DL Accelerators.}} We present the new security problem of DL accelerators on edge devices, i.e., being vulnerable to physical access based attacks. We observe that memory encryption that has been efficiently used in CPU systems causes significant (up to 50\%) performance degradation when being directly used in DL accelerators. By analyzing experimental results, we present the insights that the big bandwidth gap between the GDDR bus and the encryption engine is the main reason of causing performance degradation.

  \textbf{\emph{$\bullet$ Criticality-aware Smart Encryption for NN Models.}} We propose a criticality-aware smart encryption (SE) scheme to allow partial data to bypass the encryption engine for improving the data access bandwidth in DL accelerators without any loss of security. The idea of the SE scheme is to measure the relative importance of weight parameters in the NN model. Based on the relative importance, the SE scheme does not encrypt these weight parameters with the lowest importance, and thus it is unnecessary to encrypt their corresponding channels in the input or output feature maps. \draftsnew{Based on the quantitative security evaluation \postmicro{in terms of both IP protection and adversarial attacks}~\cite{goodfellow2014generative,kurakin2016adversarial,goodfellow2018making,Papernot2017PBA3052973}, we determine the percentage of encrypted data with which the SE scheme achieves the same security level as the full encryption scheme.}

  \textbf{\emph{$\bullet$ Colocation Mode Encryption for DL Accelerators.}} In order to improve the efficiency of memory encryption, we propose a colocation mode encryption (ColoE) scheme to store the data and its counter in the same memory line, unlike the traditional counter mode encryption storing them separately. Thus the ColoE scheme removes extra memory accesses from counters to improve the system performance and does not need a large on-chip counter cache compared with the traditional counter mode encryption. Due to the usage of counters for encryption, the ColoE scheme also has higher security level than traditional direct encryption.

  \textbf{\emph{$\bullet$ Implementation and Evaluation.}} We have implemented SEAL in GPGPU-Sim~\cite{bakhoda2009analyzing} and evaluated it using three classical CNN models including VGG-16~\cite{simonyan2014very}, ResNet-18~\cite{he2016deep}, and ResNet-34~\cite{he2016deep}. Experimental results show that, compared with traditional direct and counter mode encryption, SEAL achieves $1.4\times-1.6 \times$ IPC improvement \postmicro{and $39\%-60\%$ of latency reduction}. Compared with a baseline accelerator \draftsnew{without memory encryption}, SEAL is able to improve the security with a slight overhead ( $5\%-7\%$ IPC).

%% file: background.tex
\vspace{-4px}
\section{Background and Motivation}
\vspace{-2px}

\subsection{Deep Learning Accelerators}
Deep learning (DL)~\cite{lecun2015deep} is widely used in current artificial intelligence applications, such as natural language processing, speech recognition, and computer vision. Achieving high accuracy and low processing latency in these applications requires complicated deep learning computation~\cite{simonyan2014very,he2016deep}. Therefore, various DL hardware accelerators~\cite{jouppi2017datacenter,chen2017eyeriss,chen2014diannao,chen2014dadiannao,Nvidia2017GTX} are used to deliver high performance.

GPU is the most widely used DL accelerator due to compatibility with different algorithms and high parallelism.
The powerful parallel processing ability of GPU is efficient and suitable for DL with a large amount of parallel floating-point and matrix/vector multiplication computation.
FPGA is an alternative for implementing DL accelerators with energy efficiency. Furthermore, various ASIC DL accelerators are proposed to speed up special machine learning algorithms, such as TPU~\cite{jouppi2017datacenter}, DianNao family~\cite{chen2014diannao,chen2014dadiannao}, and Eyeriss~\cite{chen2017eyeriss}.

\begin{figure}[t]
    \vspace{-4px}
  \centering
    \includegraphics [width=0.36\textwidth]{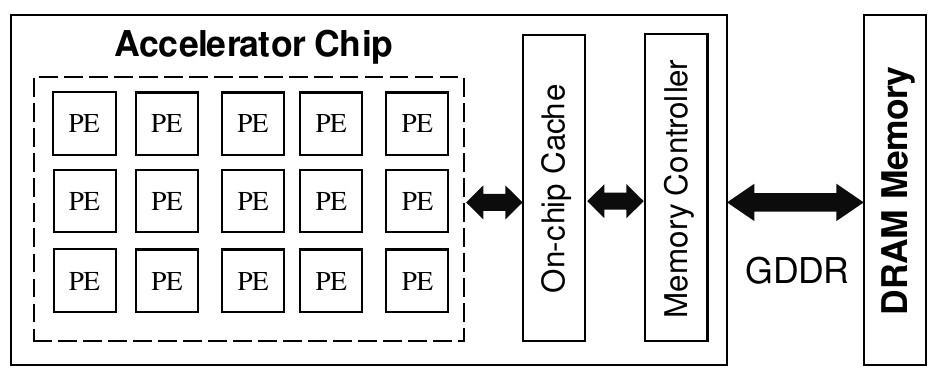}
    \vspace{-5px}
    \caption{\label{fig:1-accelerator-architecture} \textbf{The generic DL accelerator architecture.}}
    \vspace{-14px}
\end{figure}

A generic hardware architecture for these GPU, FPGA, and ASIC DL accelerators is shown in Figure~\ref{fig:1-accelerator-architecture}. The accelerator architecture consists of an array of processing elements (PEs, or called cores in GPUs) and a data cache (or called global buffer) on chip. Each PE has its own control logic and scratchpad, and communicates with the data cache through network-on-chips (NoCs). As the size of the on-chip data cache is limited, the entire NN model and the intermediate data produced during DL inference are stored in the off-chip DRAM memory with large capacity. The accelerator accesses the DRAM through the high-bandwidth GDDR bus. 

\vspace{-3px}

\vspace{-3px}
\subsection{Threat Model and Purposes}
\label{background-adversarial-attacks}

For DL applications, neural network (NN) models are critical data maintained in DL accelerators~\cite{hua2018reverse,hu2019neural}.
However, DL accelerators deployed on edge devices have the risk of leaking their NN models due to being vulnerable to physical access based attacks. Compared with devices deployed in the cloud, edge devices are easier to be physically accessed. For example, a user can dismantle his/her own self-driving car to look into the internal computer system. Therefore, DL accelerators on edge devices have the security vulnerability to physical access based attacks, i.e., bus snooping~\cite{Yan2006ICP,shi2004architectural}.

\textbf{Threat Model:} \iscare{Like existing threat models for hardware attacks on CPUs~\cite{Yan2006ICP,shi2004architectural} and accelerators~\cite{hua2018reverse,hu2019neural}, we consider on-chip components of accelerators and DRAM are secure.} However, an adversary can insert a bus snooper or a memory scanner on the GDDR memory bus\footnote{\isca{As we aim to protect the confidentiality of NN models, bus tampering attacks are not considered in our threat model that can be defended via Merkle Trees based authentication techniques~\cite{suh2003efficient}, which are orthogonal to our work.}} to obtain the data communicated between the accelerator chip and off-chip DRAM~\cite{hua2018reverse,hu2019neural}, and further steals the entire NN model.

\textbf{Threat Purposes:} We consider two threat purposes that an adversary obtains NN models via bus snooping.

\emph{1) IP Strealing.} NN models are considered as the Intellectual Property (IP) of  model owners~\cite{riazi2018privacy,hua2018reverse,tramer2016stealing}.  Model owners may consume a large amount of financial and material resources to train a sophisticated NN model. The adversary may be a business competitor of model owners. The leakage of NN models incurs the property loss of model owners and reduces their competitive advantages.

\emph{2)  Adversarial Attacks.} The exposion of an NN model can significantly increase the risk that the NN model is attacked by adversarial attacks. In adversarial attacks, an adversary aims to apply an imperceptible non-random perturbation on the input data to change the prediction results of NN models~\cite{szegedy2014intriguing,goodfellow2014generative}. The perturbed input data are termed as adversarial examples. If the adversary does not know the NN model, the adversarial attack is called \emph{black-box} attack. If the adversary knows the entire NN model, the adversarial attack is called \emph{white-box} attack. In the black-box attacks, the attack success rate is low. In the white-box attacks, the attack success rate significantly increases since the adversary can generate high-quality adversarial examples by using the known model information~\cite{liu2017delving,oh2018towards}.

In order to protect the NN models in DL accelerators from bus snooping attacks, encrypting the data transmitted through the GDDR bus is important. Existing memory encryption techniques~\cite{Yan2006ICP,shi2004architectural} are widely used in secure CPU systems to enable secure data transmission through the DDR bus of CPU memory. However, data security on the GDDR memory bus for DL accelerators are rarely touched by existing work. In this following, we first present memory encryption techniques for secure CPUs ($\S$\ref{sec:background-encryption}) and then investigate whether the straightforward solutions that perform CPU memory encryption directly on DL accelerators are efficient ($\S$\ref{sec:background-straightforward-solution}).

\begin{figure}[t]
    \vspace{-6px}
  \centering
  \subfloat[Direct Encryption]{
    \label{fig:background-direct-encryption}
    \includegraphics[width=0.42\textwidth]{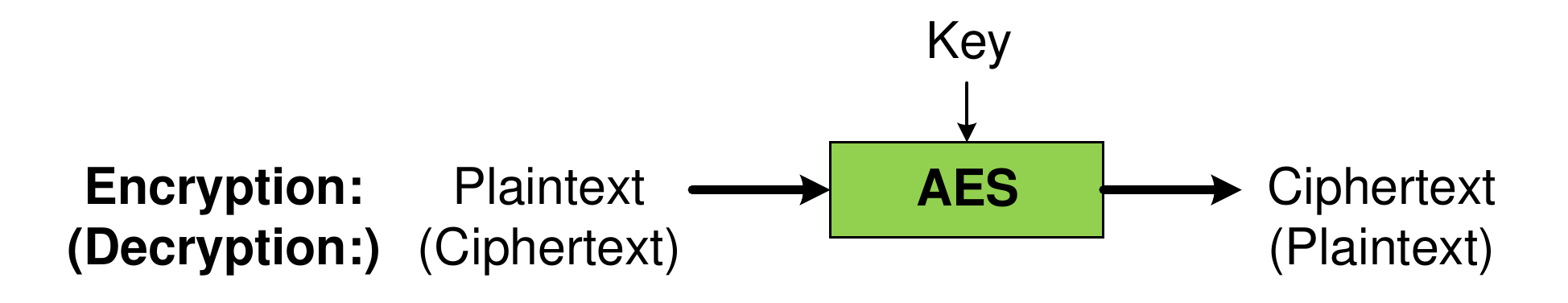}
    }

    \vspace{-8px}
    \subfloat[Counter mode encryption]{
    \label{fig:background-counter-mode-encryption}
    \includegraphics[width=0.42\textwidth]{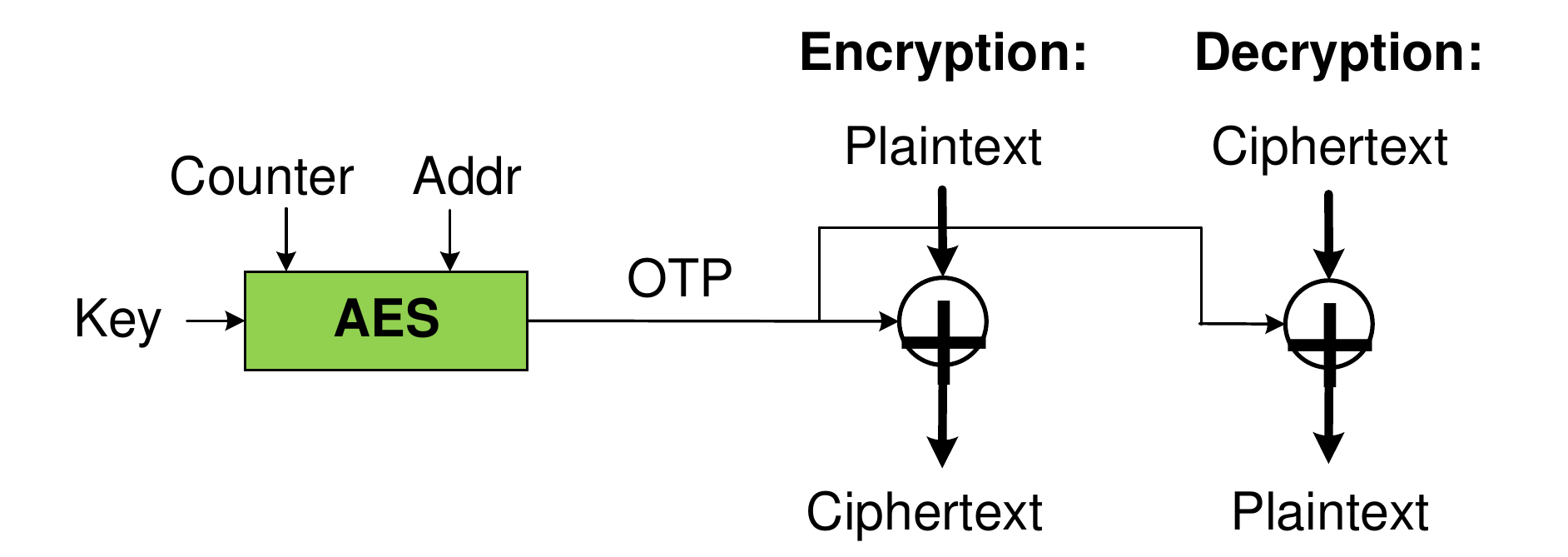}}
    \vspace{-4px}
    \caption{\textbf{Encryption and decryption operations in the direct encryption and counter mode encryption.}}
    \vspace{-16px}
\end{figure}

\vspace{-3px}
\subsection{Memory Encryption}
\label{sec:background-encryption}
In secure CPUs, the encryption engine of a block cipher algorithm (e.g., AES~\cite{daemen2013design}) is added in the memory controller for encrypting and decrypting data. In general, there are two memory encryption models used for secure CPUs, including direct encryption and counter mode encryption.

As shown in Figure~\ref{fig:background-direct-encryption}, in the direct encryption, each cache line is encrypted by the AES encryption engine before being written back to the DRAM memory. After being read from the DRAM memory, each line is decrypted and then put into the last level cache. However, direct encryption causes high decryption latency in the critical path of memory accesses in CPUs. Additionally, direct encryption encrypts all memory lines by using the same global key, which has a low security level. \draftsnew{Since the same data are always encrypted to the same ciphertext, it leaves direct encryption vulnerable to dictionary and retry based attacks}~\cite{Young2015DWE,awad2016silent}.

As shown in Figure~\ref{fig:background-counter-mode-encryption}, in counter mode encryption~\cite{lipmaa2000ctr}, a global key, the line address and the per-line counter pass through the AES encryption engine to generate a one-time pad (OTP). The plaintext or ciphertext is then encrypted or decrypted by simply XORing its OTP. Each memory line in the off-chip DRAM has a counter. All counters are stored in the DRAM. Recently used counters are buffered in an on-chip counter cache managed by the memory controller. If the counter of a memory line to be read is in the counter cache, its decryption latency is hidden in the memory read latency, since the OTP is generated in parallel with the memory read. Only the XOR latency is added to the critical path, thus reducing the decryption latency.

Moreover, counter mode encryption provides a higher security level than direct encryption, since OTPs are never reused for data encryption which keep counter mode encryption secure from dictionary and retry based attacks. First, since the line address is used to generate the OTP, the data stored at different addresses are encrypted by different OTPs. Second, a per-line counter is used to generate the OTP and the counter increases one on each write. Data rewritten in the same address are encrypted by different OTPs.
In general, counters are stored in the plaintext since data cannot be decrypted if an adversary has the knowledge of the counter value but does not know the key~\cite{lipmaa2000ctr,Young2015DWE}.

\vspace{-3px}
\subsection{Straightforward Solutions for Securing DL Accelerators}
\label{sec:background-straightforward-solution}

We consider two straightforward solutions, i.e., simply employing the direct encryption and counter mode encryption in DL accelerators, to improve the security of NN models. Without loss of generality, in the rest of this paper, we analyze GPU as a representative example of DL accelerators. However, the problems, insights, and solutions that we develop are also applicable to other DL accelerators.

\begin{figure}[t]
  \vspace{-15px}
  \centering
  \subfloat[Instruction per cycle (IPC)]{
    \label{fig:background-motivation-1}
    \includegraphics[width=0.24\textwidth]{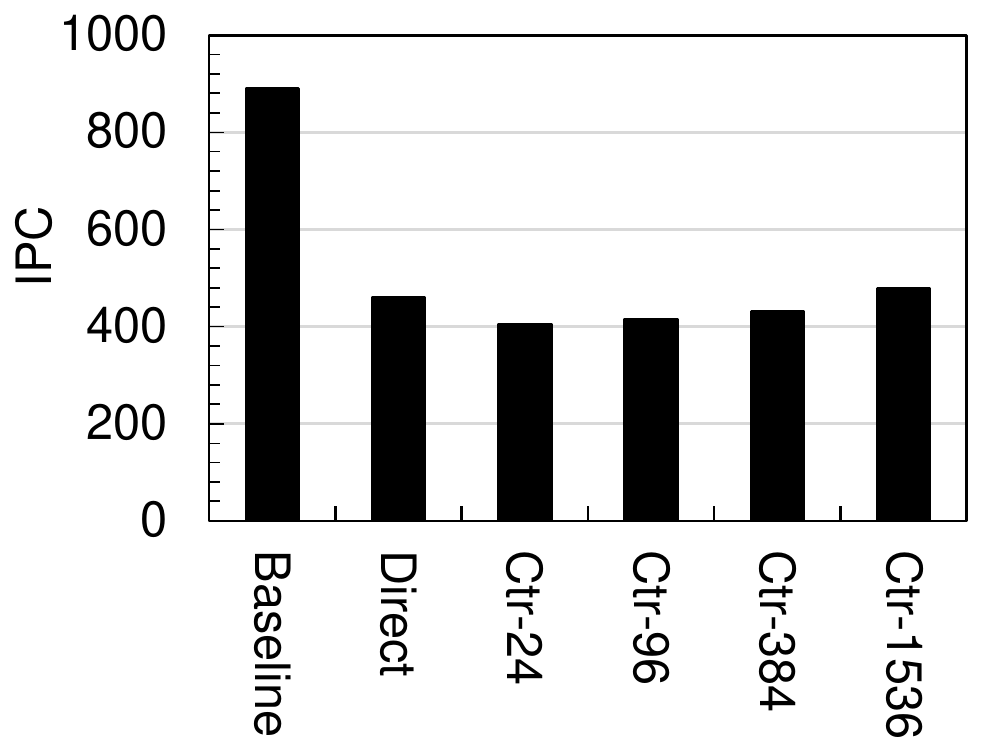}
    }
    \subfloat[Counter cache hit rate]{
    \label{fig:background-motivation-2}
    \includegraphics[width=0.24\textwidth]{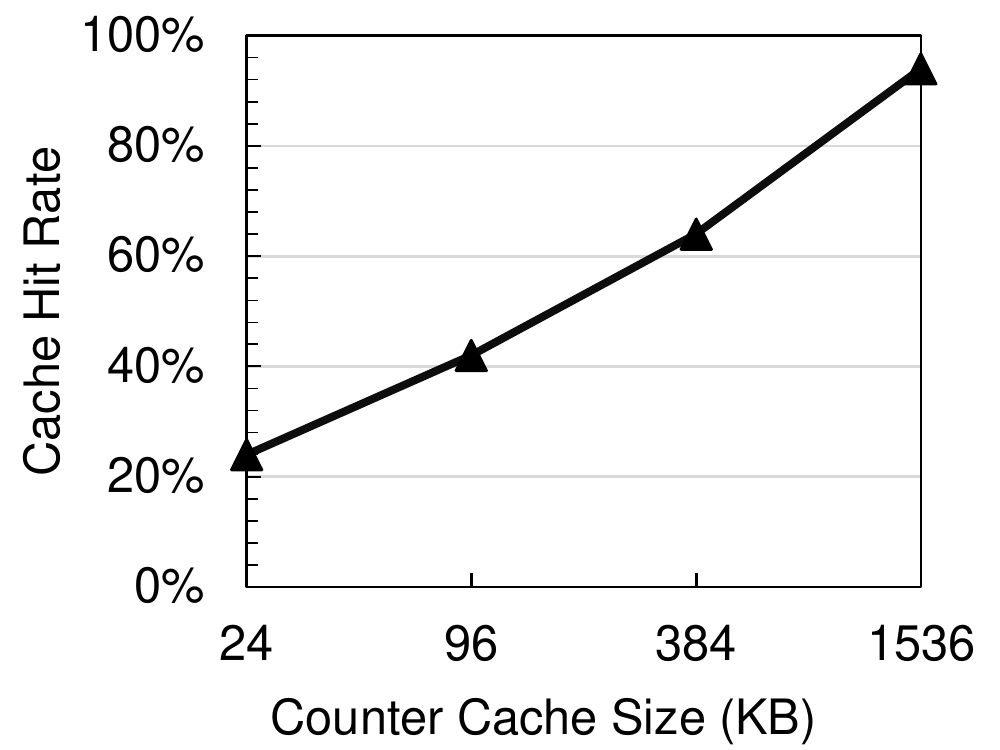}}
    \vspace{-5px}
    \caption{\textbf{The IPC of GPUs with two straightforward memory encryption solutions.} \textit{("Baseline" means a baseline GPU without using memory encryption. "Direct" means the direct encryption. "Ctr-96" means the counter mode encryption with the 96KB counter cache and each memory controller has a 16KB (=96KB/6) counter cache.)}}
    \vspace{-20px}
\end{figure}

We implement the two straightforward solutions in GPGPU-Sim~\cite{bakhoda2009analyzing}. Since the encryption engine increases the chip area and energy overhead that also affects the chip cooling~\cite{Awad2017OLA,opencores2012aes,mathew201053gbps}, each memory controller generally includes one encryption engine~\cite{Yan2006ICP,awad2016silent,liu2018crash,Young2015DWE}. \isca{Thus the six memory controllers in the modeled GPU include six encryption engines.} For the counter mode encryption, we add an on-chip counter cache to buffer recently used counters. The detailed GPU configurations are shown in Section~\ref{sec:evaluation-methodology-config}. We use the modeled GPU to execute matrix multiplication computation that is the most common operation in DL algorithms. We evaluate the IPC (instruction per cycle) of the GPU with different encryption schemes and compare them with a baseline GPU without using memory encryption, as shown in Figure~\ref{fig:background-motivation-1}.

First, \draftsnew{we observe the GPU with memory encryption is significantly less efficient than the GPU without memory encryption.} Memory encryption decreases the GPU IPC by $45\%-54\%$ for the matrix multiplication computation. 
Second, using the counter mode encryption scheme does not deliver higher performance compared to using the direct encryption scheme on GPU. With the small counter cache sizes, i.e., 24KB, 96KB, and 384KB, the performance of the counter mode encryption scheme is even lower than the direct encryption scheme. \isca{By using a large counter cache, i.e., 1536KB, the IPC of the GPU is improved by 15\%. However, the counter cache size is double of the L2 cache size in the modeled GPU as shown in the configurations (Section~\ref{sec:evaluation-methodology-config}), which is too large to be deployed on the GPU die.}

The reason that memory encryption significantly reduces the GPU performance is the big bandwidth gap between the GDDR memory bus and the encryption engine, as shown in Table~\ref{tab:backgorund-banwidth}. In CPU systems, memory encryption works well~\cite{Yan2006ICP,shi2004architectural} since the AES encryption engine has a similar bandwidth to the DDR memory bus and the PCIe bus of CPU. However, in GPU systems, the GPU performance is highly bandwidth-bounded and hence the GDDR memory is designed for GPUs to achieve high memory access bandwidth. The bandwidth of the GDDR memory bus is generally more than 160GB/s~\cite{Nvidia2018Turing,Nvidia2014GTX980,Nvidia2017GTX,Nvidia2012GTX480}. However, the state-of-the-art \isca{pipelined} AES encryption engine with hardware implementation achieves only about 8GB/s of bandwidth on average~\cite{mathew201053gbps}. \isca{Even though we deploy one encryption engine in every memory controller, the total encryption bandwidth is 48 GB/s. As a result, the high bandwidth of the GDDR memory bus is under-utilized and the AES encryption engine becomes the bandwidth bottleneck in secure GPUs.
\iscare{A single AES engine usually occupies over 1 $mm^2$ on-die area and has hundreds or thousands of mW power, as shown in Table~\ref{tab:AES-engines}. As resources on the microprocessor die are very scarce, it is ruinously costly to integrate more encryption engines into memory controllers on the GPU die~\cite{cryptoeprint2016204}. Even though a GPU/CPU die usually has an area of $90-600$ $mm^2$, most area is occupied by cores and on-die memory and only less than 10\% area is left to memory controllers~\cite{Reddit2019,Khalid2016}. This is also the reason why Intel carefully designs the AES hardware implementation to reduce area and energy overheads for Software Guard Extensions (SGX)~\cite{cryptoeprint2016204}.
Like the design principle of Intel's SGX~\cite{cryptoeprint2016204} and many previous works~\cite{Yan2006ICP,awad2016silent,liu2018crash,Young2015DWE}, the goal of this paper is also to improve the hardware security while having low on-die overheads.}
Moreover, since the data access bandwidth is the performance bottleneck,
the counter mode encryption incurs extra memory access requests for reading and writing counters compared with the direct encryption, thus delivering low performance with small cache sizes}.

\begin{table}[t]
\caption{\textbf{Bandwidth comparisons of AES encryption engine and different buses}~\cite{hennessy2011computer,mathew201053gbps,Nvidia2018Turing}.}
  \vspace{-8px}
\label{tab:backgorund-banwidth}
\footnotesize
\begin{center}
   \begin{tabular}{|c|l|c|}
    \hline
    \multirow{2}{*}{DDR bus} & DDR3 (No.800$-$2666) & $6.4 \sim 21.3$ GB/s \\
    \cline{2-3}
    & DDR4 (No.1600$-$3200) & $12.8 \sim 25.6$ GB/s \\
    \hline
    \multirow{2}{*}{PCIe bus} & PCIe 3.0 ($\times8$ links) & 8 GB/s \\
    \cline{2-3}
    & PCIe 3.0 ($\times16$ links) & 16 GB/s \\
    \hline
    AES engine & 128bit block & $1.5 \sim 19$ GB/s \\
    \hline
    \multirow{2}{*}{GDDR bus} & GDDR5 & $160 \sim 336$ GB/s \\
    \cline{2-3}
    & GDDR5X & $320 \sim 484$ GB/s \\
    \hline
\end{tabular}
\end{center}
\vspace{-10px}
\end{table}

\begin{table}[t]
\caption{\textbf{\iscare{Performance comparisons of different AES encryption engine implementations (counter mode)}}.}
  \vspace{-14px}
\label{tab:AES-engines}
\footnotesize
\begin{center}
   \begin{tabular}{|c|p{0.72cm}<{\centering}|p{0.9cm}<{\centering}|p{1cm}<{\centering}|p{1.34cm}<{\centering}|}
    \hline
    & Area ($mm^2$) & Power (mW) & Latency (cycle) & Throughput (GB/s) \\
    \hline
    Morioka et al.~\cite{morioka200410} & N/A & 1920  &  10    & 1.5   \\ 
    \hline
    Mathew et al.~\cite{mathew201053gbps} & 1.1 & 125  &  20  & 6.6   \\
    \hline
    Ensilica~\cite{Ensilica2020} & 1.4 & N/A  &  11  & 8   \\
    \hline
    Sayilar et al.~\cite{sayilar2014cryptoraptor} & 6.3 & 6207  &  20    &  16  \\
     \hline
    Liu et al.~\cite{liu2011parallel} & 6.6 & 1580  &  152    &  19  \\
    \hline
\end{tabular}
\end{center}
\vspace{-18px}
\end{table}

%% file: design.tex
\vspace{-2px}
\section{The SEAL Design}
\vspace{-1px}

\begin{figure*}[t]
    \vspace{-11px}
  \centering
    \includegraphics [width=0.94\textwidth]{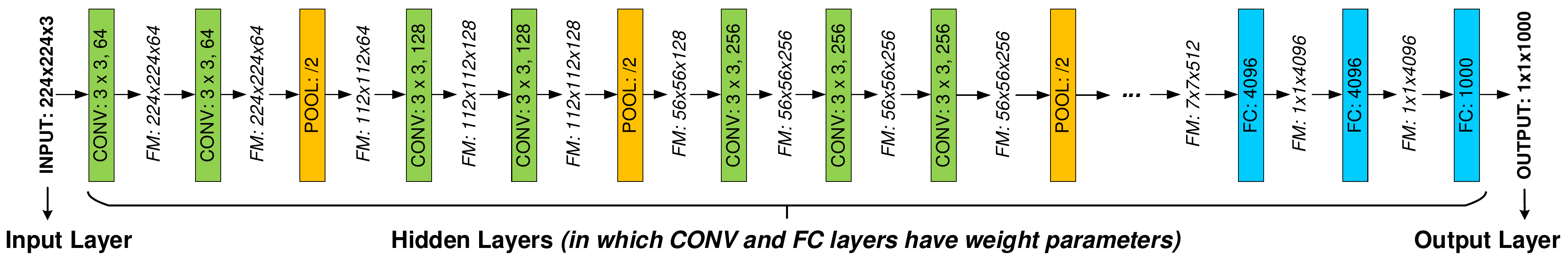}
    \vspace{-7px}
    \caption{\label{fig:design-VGG} \textbf{The CNN architecture of VGG-16 as an example.} \textit{(The CNN uses an image with $224\times224$ pixels and 3 channels as the input layer and outputs a one-dimensional vector. The input and output of each hidden layer are called feature maps (FMs) and output FMs of the previous layer are input FMs of the latter layer. (CONV: $3\times3$, 64) indicates a convolution layer with the $3\times3$ CONV kernel and 64 output channels. (FM: $224\times224\times64$) means the FMs with the size of $224\times224\times64$. $POOL$ indicates a pooling later and FC indicates a full connected layer.}}
    \vspace{-12px}
\end{figure*}

We propose SEAL, a secure and efficient DL accelerator scheme to enhance the security of NN models.
SEAL improves the performance of secure DL accelerators by exploring and exploiting software and hardware co-design. In the software layer, to improve the data access bandwidth of DL accelerators, a criticality-aware smart encryption (SE) scheme ($\S$\ref{sec:design-smart-encryption}) is used to measure the relative importance of weight parameters in the NN model. Only the relatively important weight parameters are processed by the AES encryption engine and the remaining parameters bypass the AES encryption engine, which reduces the amount of data to be encrypted without compromising the security. \postmicro{We quantitatively analyze and evaluate the security of the SE scheme in terms of both IP protection and adversarial attacks, and leverage the evaluation results to guide the parameter configuration of the SE scheme to obtain the maximum performance benefit and the highest security level ($\S$\ref{sec:security-analysis}).}  In the hardware layer, to improve the efficiency of memory encryption, SEAL leverages a colocation mode encryption (ColoE) scheme ($\S$\ref{sec:design-data-counter-colocation}) to achieve the same security level as the counter mode encryption while eliminating extra memory accesses from counters. Moreover, we present the overall hardware architecture design to support SE and ColoE  ($\S$\ref{sec:design-harware-architecture}).

\vspace{-2px}
\subsection{Criticality-aware Smart Encryption}
\label{sec:design-smart-encryption}

In this subsection, we first use the convolutional neural network (CNN) that is a widely used neural network for DL as an example to present the challenges of performing partial encryption on DL accelerators. We then present the proposed criticality-aware smart encryption scheme.

\subsubsection{Challenges for Partial Encryption}
\label{sec:design-challenges}

During the process of the CNN inference, there are four kinds of data, i.e., data in the input layer, data in the output layer, weight parameters in hidden layers (i.e., the NN model data), and intermediate data (i.e., feature maps) produced by hidden layers, as shown in Figure~\ref{fig:design-VGG}. If we encrypt all the data during the CNN inference, the inference performance significantly decreases, as presented in Section~\ref{sec:background-straightforward-solution}. This is mainly because the bandwidth of the AES encryption engine is far lower than that of the GDDR memory bus, limiting the total data access bandwidth. If we can only encrypt partial data to reduce the amount of data to be encrypted, the total data access bandwidth improves. Nevertheless, performing partial encryption is not easy due to the following fundamental challenges.

\textbf{\emph{Challenge 1: How to select appropriate data to be encrypted?}} Among the four kinds of data, the data in the input and output layers are usually known by the adversary. For example, for the DL accelerator in a self-driving car, the input data are the pictures of the current visual field taken by cameras, which can be obtained by the adversary. The output data are the current actions of the car, e.g., stop, turning left, or turning right, also known by the adversary. A simple way of the partial encryption is that we do not encrypt the data in the input and output layers and encrypt the remaining data including weight parameters in the NN model and intermediate data produced by hidden layers. However, the sizes of the data in the input and output layers are far smaller than that of the intermediate data, as shown in Figure~\ref{fig:design-VGG}. For example, the data in the input layer with the size of $224\times224\times3$ are 11 times smaller than the output feature maps of the first CONV layer with the size of $224\times224\times64$. Therefore, this simple way is inefficient to improve inference performance.

Moreover, among these data in the CNN inference, weight parameters of the NN model have to be protected. Intuitively, \draftsnew{we can encrypt only the weight parameters of the NN model and do not encrypt the remainder of the data to reduce the encryption overhead.} However, an adversary can calculate or speculate the weight parameters of the NN model via unencrypted feature maps. For example, a CONV layer computes the input feature maps $X$ with a kernel matrix $\omega$ to produce the output feature maps $Y$, i.e., $Y = X \omega$. If $X$ and $Y$ are not encrypted, an adversary can easily compute the kernel matrix $\omega$ via the equation $\omega = X^{-1} Y$ in which $X^{-1}$ is the inverse matrix of $X$. Therefore, it is important to protect the NN model data from being calculated or speculated from the unencrypted data.

\textbf{\emph{Challenge 2: How to evaluate the impact of \isca{partial} encryption on security?}} Intuitively, encrypting all data inputted and produced during the NN inference has a high security level but causes significant performance degradation. Selectively un-encrypting partial data can improve the performance which however may exacerbate security. An adversary can directly compute encrypted weights via unencrypted feature maps as discussed above. Moreover, existing fine-tuning techniques~\cite{li2016pruning,radenovic2016cnn} for NN models can also be used to speculate a complete NN model based on known partial weight parameters and the data in the input and output layers. Specifically, the adversary can fill the known partial weight parameters in the NN model and then use the data in the input and output layers to retain a complete NN model. Hence, how to evaluate and quantify the impact of partial encryption on security is non-trivial for designing an efficient encryption scheme.

\vspace{-3px}
\subsubsection{Smart Encryption in SEAL}
\label{sec:selective-encryption-scheme}

To address these challenges, we propose \emph{a criticality-aware smart encryption (SE) scheme} in SEAL, which aims to reduce the amount of encrypted data while improving the NN model security. The SE scheme quantitatively measures the relative importance of weight parameters in each layer by calculating the sum of their absolute weights, i.e., $\ell_1$-norm. \draftsnew{The weight parameters with the smallest absolute values in each layer are considered to be least important and hence are not encrypted.} Thus it is unnecessary to encrypt the corresponding channels in the input or output feature maps of unencrypted weight parameters. As a result, the amount of data to be encrypted is significantly reduced. The percentage of un-encrypted weight parameters is determined based on the quantitative security evaluation in Section~\ref{sec:security-analysis} to obtain maximum performance benefit and highest security level.

\begin{figure}[t]
	\centering
	\includegraphics[width=0.45\textwidth]{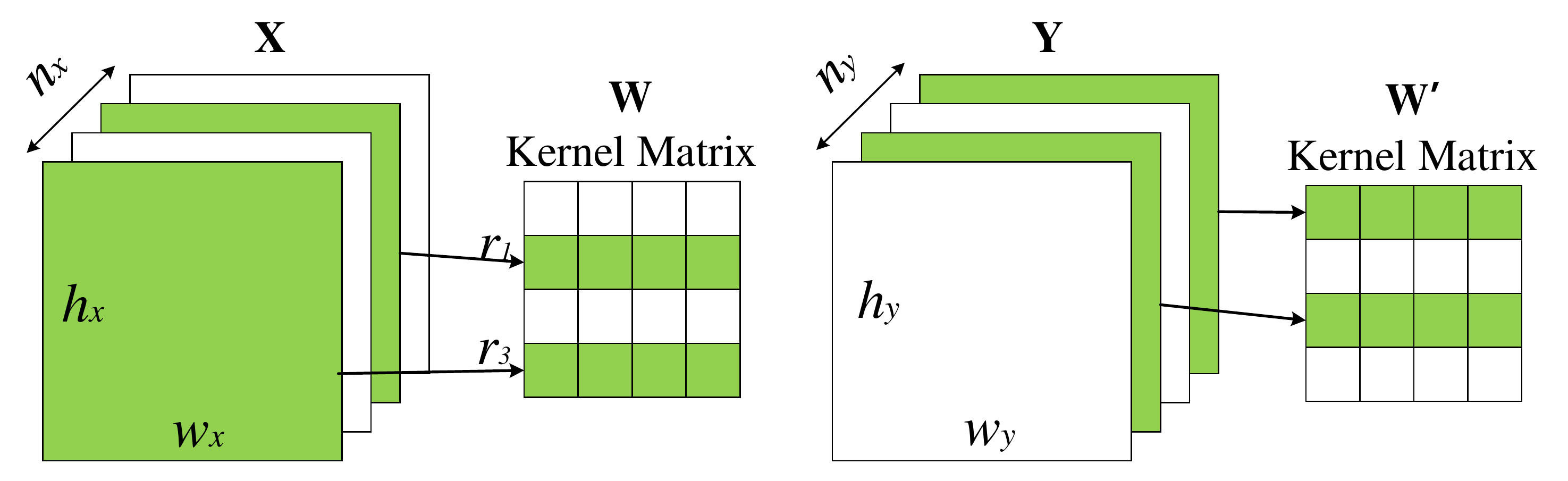}
	\vspace{-10px}
	\caption{\textbf{An example for the smart encryption scheme.} \textit{(Green areas: encrypted data. \iscare{Each grid in the kernel matrix is a kernel, e.g., 3*3.})}}
	\label{fig:design-selective-encryption}
	\vspace{-14px}
\end{figure}

In deep neural networks, we consider use the SE scheme in the CONV layers since most layers in a CNN model are CONV layers, e.g., 13/16 for VGG-16, 17/18 for ResNet-18, and 33/34 for ResNet-34. The computation process of a CONV layer is shown in Figure~\ref{fig:design-selective-encryption}. Weight parameters in a CONV layer are organized as a convolutional kernel matrix, and each convolutional kernel is a weight matrix, e.g., $3\times3$. \draftsnew{The computation of a CONV layer transforms the input feature maps with the convolutional kernel matrix to the output feature maps.} The convolutional kernel matrix has $n_x$ kernel rows and $n_y$ kernel columns. $n_x$ is equal to the number of channels in the input feature maps. Each kernel row in the kernel matrix corresponds to a single input channel in the input feature maps and this input channel does not involve the convolution computation with other kernel rows, as shown in Figure~\ref{fig:design-selective-encryption}. Similarly, $n_y$ is equal to the number of channels in the output feature maps. Each kernel column in the kernel matrix corresponds to a single output channel in the output feature maps, as shown in Figure~\ref{fig:design-selective-encryption}.

\textbf{Relative Importance Measurement.} We first present our approach for relative importance measurement as shown in Figure~\ref{fig:design-selective-encryption}. We measure the relative importance of a kernel row in each layer by calculating the sum of its absolute weights, i.e., $\ell_1$-norm. The sum of absolute weights in a row also represents the average magnitude of the kernel weights which gives an expectation of the magnitude of the output feature map. Thus kernel rows with smaller sums of absolute weights tend to produce feature maps with weak activations, compared with the other kernel rows in the same layer~\cite{li2016pruning}. Hence, these rows with small absolute-value sums have a lower impact on the output of the entire NN model compared with the rows with large absolute-value sums. Existing work~\cite{li2016pruning,han2015deep} on pruning NN models demonstrate that, even after completely eliminating the convolution computation that uses these weight parameters with small absolute values, the original accuracy of the NN model can be regained by retraining the networks. This observation indicates that these weight parameters with small absolute values are less important to the NN model and thus rarely affect the security of the NN model. We have confirmed this conjecture by performing \postmicro{IP protection and} adversarial attack tests as presented in Section~\ref{sec:security-analysis}, whose results motivate us to propose the smart encryption (SE) scheme to reduce the encryption overhead in DL accelerators by only encrypting the weight parameters with large absolute values.

\textbf{Smart Encryption (SE).} After computing the sum of absolute weights in each row, the SE scheme sorts the kernel rows based on their sums. The SE scheme then encrypts partial kernel rows with the largest sums. The percentage of the encrypted kernel rows is determined by our quantitative security analysis as shown in Section~\ref{sec:security-analysis}. However, the encrypted weight parameters in the SE scheme can be figured out if the input and output feature maps of this CONV layer are unencrypted as discussed in Section~\ref{sec:design-challenges}. Therefore, for each encrypted row, the SE scheme also encrypts one input channel in the input feature maps corresponding to the encrypted row, since each kernel row corresponds to a single input channel and does not involve the convolution computation with other input channels, as shown in Figure~\ref{fig:design-selective-encryption}. In this way, the encrypted weight parameters cannot be figured out. For example, for the matrix multiplication $Y = X \omega$, the input channel $X$ and the weights $\omega$ are encrypted. $\omega$ cannot be figured out even though the adversary knows $Y$.
\postmicro{The data in the input channel $X$ is encrypted once being produced by the previous CONV layer. Hence, the plaintext in the encrypted channel $X$ is never exposed to the memory bus.}

\iscare{Moreover, when considering unencrypted data among multiple layers, the encrypted channels and weights cannot be figured out and hence also secure. To prove this, we use a simple example with two sequential CONV layers, i.e., $Y = X \omega$ and $Z = Y \omega^{'}$, as follows.}
\vspace{-4px}
\begin{equation}
\small
\centering
\label{eq:parameters}
\begin{aligned}
X = \left[
\begin{array}{cc}
\bm{X_0} & X_1 \\
\end{array}
\right],
\omega = \left[
\begin{array}{c}
\bm{\omega_{\bm{r0}}} \\
 \omega_{r1}
\end{array}
\right]
= \left[
\begin{array}{cc}
\bm{\omega_{\bm{00}}} & \bm{\omega_{\bm{01}}} \\
\omega_{10} & \omega_{11} \\
\end{array}
\right],
Y = \left[
\begin{array}{cc}
Y_0 &\bm{Y_1} \\
\end{array}
\right],
\\
\omega^{'}  = \left[
\begin{array}{c}
\omega^{'}_{r0} \\
\bm{\omega^{'}_{\bm{r1}}} \\
\end{array}
\right]
= \left[
\begin{array}{cc}
\omega^{'}_{00} & \omega^{'}_{01} \\
\bm{\omega^{'}_{\bm{10}}} & \bm{\omega^{'}_{\bm{11}}} \\
\end{array}
\right],
Z = \left[
\begin{array}{cc}
\bm{Z_0} & Z_1 \\
\end{array}
\right]
\end{aligned}
\end{equation}
\iscare{The feature maps X, Y, and Z have 2 channels. Since there are 2 input and output channels, the kernel matrixes $\omega$ and $\omega^{'}$ have 2 rows and 2 columns. With a 50\% encryption ratio, we assume the first row $\omega_{r0}$ in $\omega$ is encrypted, and  the second row $\omega^{'}_{r1}$ in $\omega^{'}$ is encrypted. Based on the SE scheme, we should encrypt the first channel $X_0$ in $X$ and the second channel $Y_1$ in $Y$. Moreover, we assume $Z_0$ is encrypted in $Z$. In Equation~\ref{eq:parameters}, the bold fonts mean encrypted data. Thus for the two sequential CONV layers, we can have the following equations (the encrypted data are in bold):}
\begin{equation}
\small
\label{eq:muti-layers-0}
\left\{
\begin{array}{lr}
\bm{X_0}*\bm{\omega_{\bm{00}}} + X_1*\omega_{10} = Y_0 \\
\bm{X_0}*\bm{\omega_{\bm{01}}} + X_1*\omega_{11} = \bm{Y_1}
\end{array}
\right.
\end{equation}
\begin{equation}
\small
\label{eq:muti-layers-1}
\left\{
\begin{array}{lr}
Y_0*\omega^{'}_{00}  + \bm{Y_1}*\bm{\omega^{'}_{\bm{10}}} = \bm{Z_0} \\
Y_0*\omega^{'}_{01}  + \bm{Y_1}*\bm{\omega^{'}_{\bm{11}}} = Z_1 \\
\end{array}
\right.
\end{equation}

\iscare{As shown in Equations~\ref{eq:muti-layers-0} and ~\ref{eq:muti-layers-1}, encrypted input channels are never multiplied with unencrypted weight rows, and unencrypted input channels are never multiplied with encrypted weight rows. Thus we can only obtain the product of two encrypted matrixes, e.g., $X_0*\omega_{00}$, but cannot figure out any single encrypted matrix from Equations~\ref{eq:muti-layers-0} and ~\ref{eq:muti-layers-1}. Therefore, the data in encrypted channels and weights are secure even considering data among multiple layers.}

\draftsnew{In fact, the SE scheme can also be applied to FC layers since each FC layer includes a kernel matrix like the CONV layer. Therefore, the proposed SE scheme can be applied to other deep neural networks, e.g., recurrent neural networks~\cite{cho2014learning,hochreiter1997long}, that are composed of many FC layers.}

\vspace{-2px}
\subsection{Colocation Mode Encryption}
\label{sec:design-data-counter-colocation}
\vspace{-1px}
There are two existing memory encryption models, i.e., direct encryption and counter mode encryption, as discussed in Section~\ref{sec:background-encryption}. Direct encryption has a lower security level due to being vulnerable to the directory and retry based attacks. Counter mode encryption enhances security by using counters for encryption but requires a large counter cache on chip to achieve a high cache hit rate. Based on previous works~\cite{Yan2006ICP,awad2016silent,liu2018crash} on counter mode encryption, the size of the used counter cache is usually up to 1MB$-$4MB. It is reasonable to add a large counter cache on CPU chips, since a large part of the area on CPU chips is occupied by memories, e.g., last level cache, and thus it is easy to partition some memories for the counter cache. However, for DL accelerators, especially GPUs, a large part of the on-chip area is used for computing units. The L2 cache for current commercial GPUs~\cite{Nvidia2014GTX980,Nvidia2017GTX,AMD2012GCN} is usually no more than several MB. Therefore, adding a large counter cache on GPU chips is unpractical.

\begin{figure}[t]
  \vspace{-14px}
  \centering
  \subfloat[Counter mode encryption]{
    \label{fig:3-6-CTR}
    \includegraphics[width=0.22\textwidth]{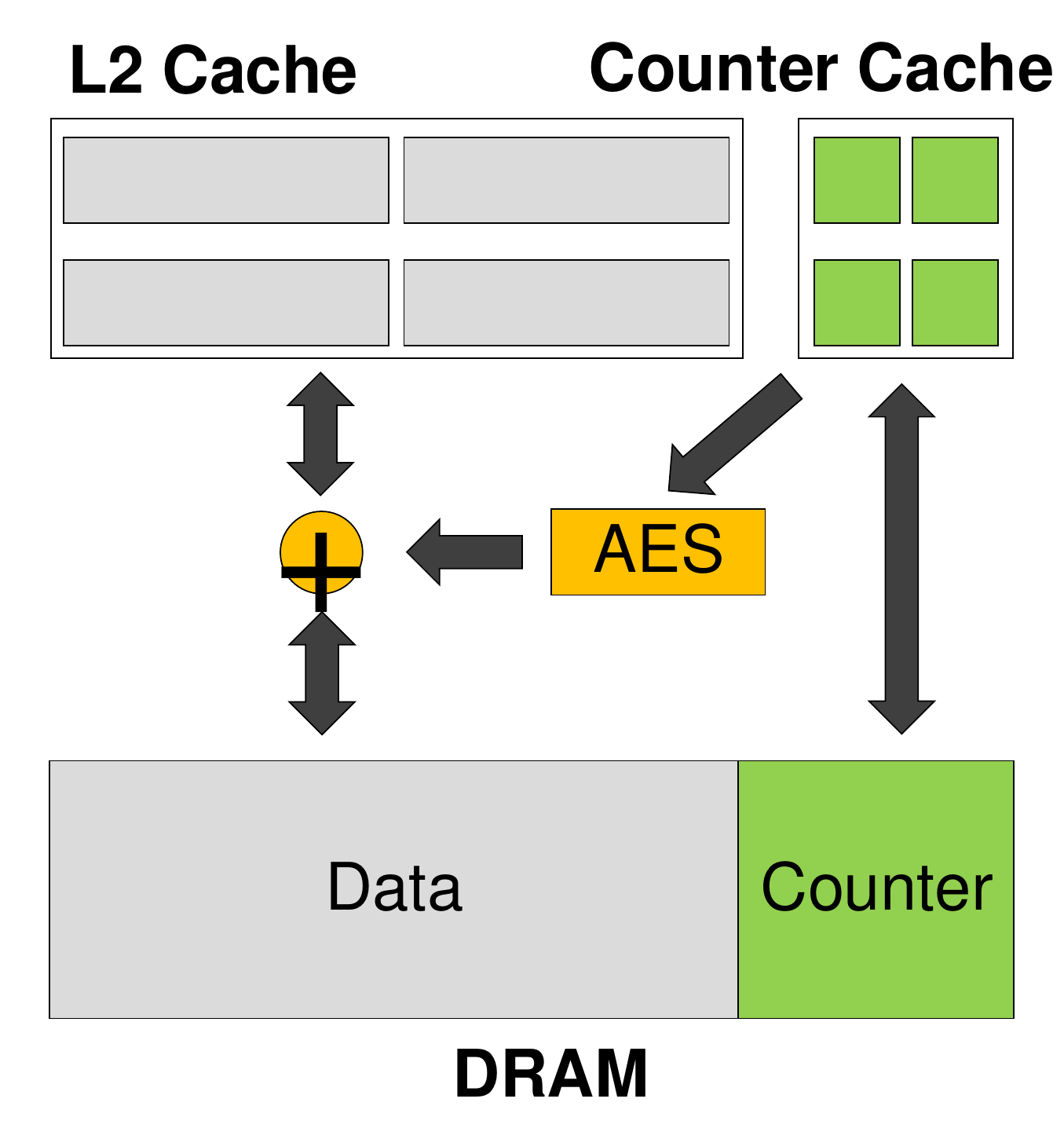}}
    \hspace{6px}
    \subfloat[Colocation mode encryption]{
    \label{fig:3-6-ColoE}
    \includegraphics[width=0.22\textwidth]{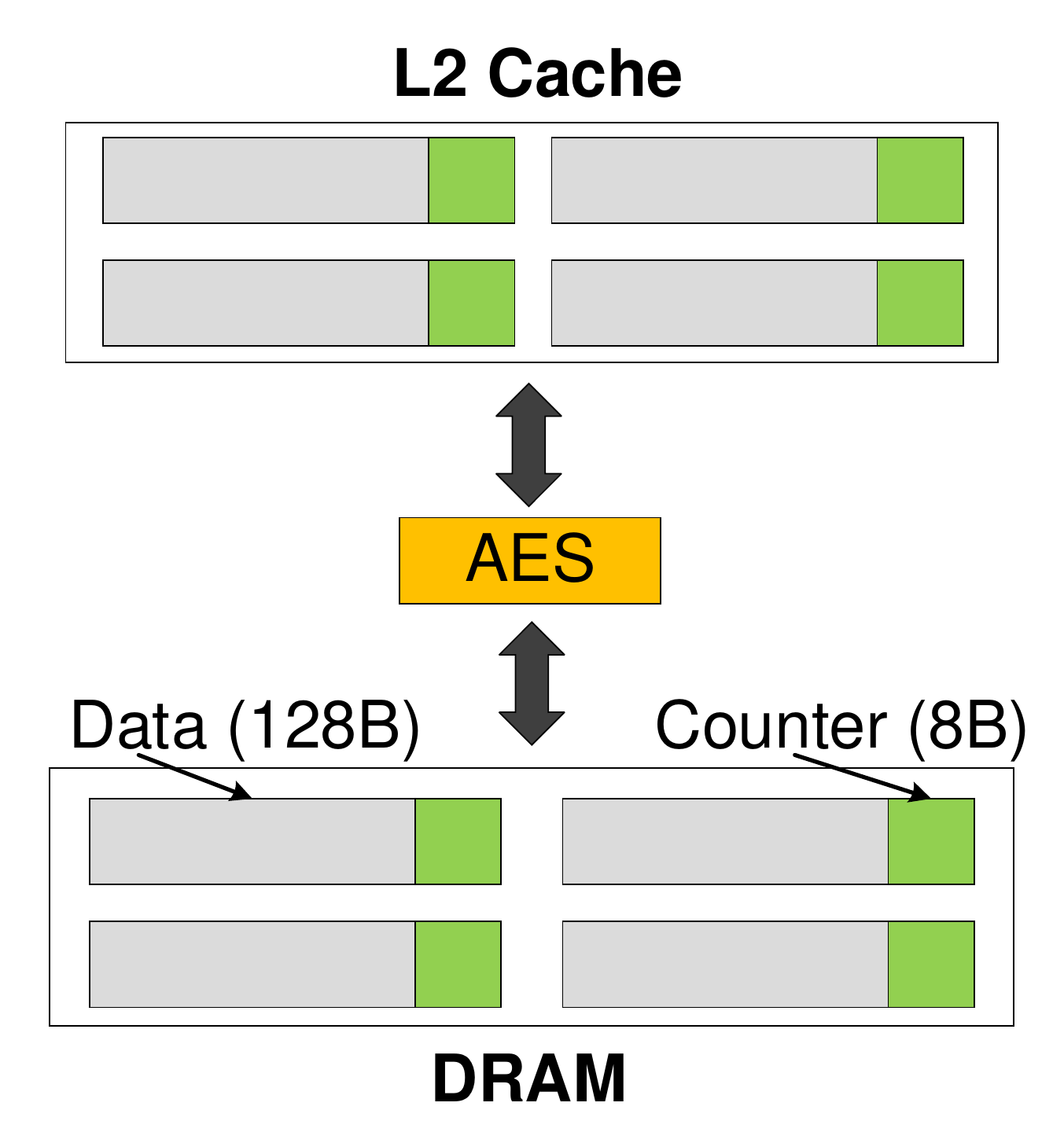}}
    \vspace{-5px}
    \caption{\textbf{The comparisons between counter and colocation mode encryption schemes.}}
    \vspace{-15px}
\end{figure}

Our paper proposes \emph{a colocation mode encryption (ColoE) scheme} for DL accelerators without using an on-chip counter cache. The ColoE scheme achieves the same security level while having higher performance on DL accelerators, compared with the traditional counter mode encryption. Unlike the traditional counter mode encryption that stores the data and their counters separately as shown in Figure~\ref{fig:3-6-CTR}, the ColoE scheme stores the data and its counter together, i.e., colocation. \isca{Like Intel's SGX~\cite{cryptoeprint2016204,gueron2016memory}, we use the monolithic counter scheme rather than the split counter scheme~\cite{Yan2006ICP} to avoid the overheads of intricate page re-encryption. The counter area is 8B for each memory line.} Thus a memory line for storing the encrypted data is 136B including 128B data and 8B counter area as shown in Figure~\ref{fig:3-6-ColoE}. When the data is evicted from the L2 cache, the ColoE scheme encrypts the data using its co-located counter, its memory address, and a key and then writes it into the DRAM memory. When the data is read from the DRAM memory, the ColoE scheme decrypts the data and then sends it to the L2 cache.
\isca{Unlike the traditional counter mode encryption that needs extra memory accesses to read/write counters, the ColoE scheme avoids these memory requests from counters by co-locating the data and their counters, thus improving the encryption performance in GPUs.}


\vspace{-3px}
\subsection{Implementation and Overall Architecture}
\label{sec:design-harware-architecture}
\begin{figure}[t]
  \vspace{-3px}
  \centering
    \includegraphics [width=0.42\textwidth]{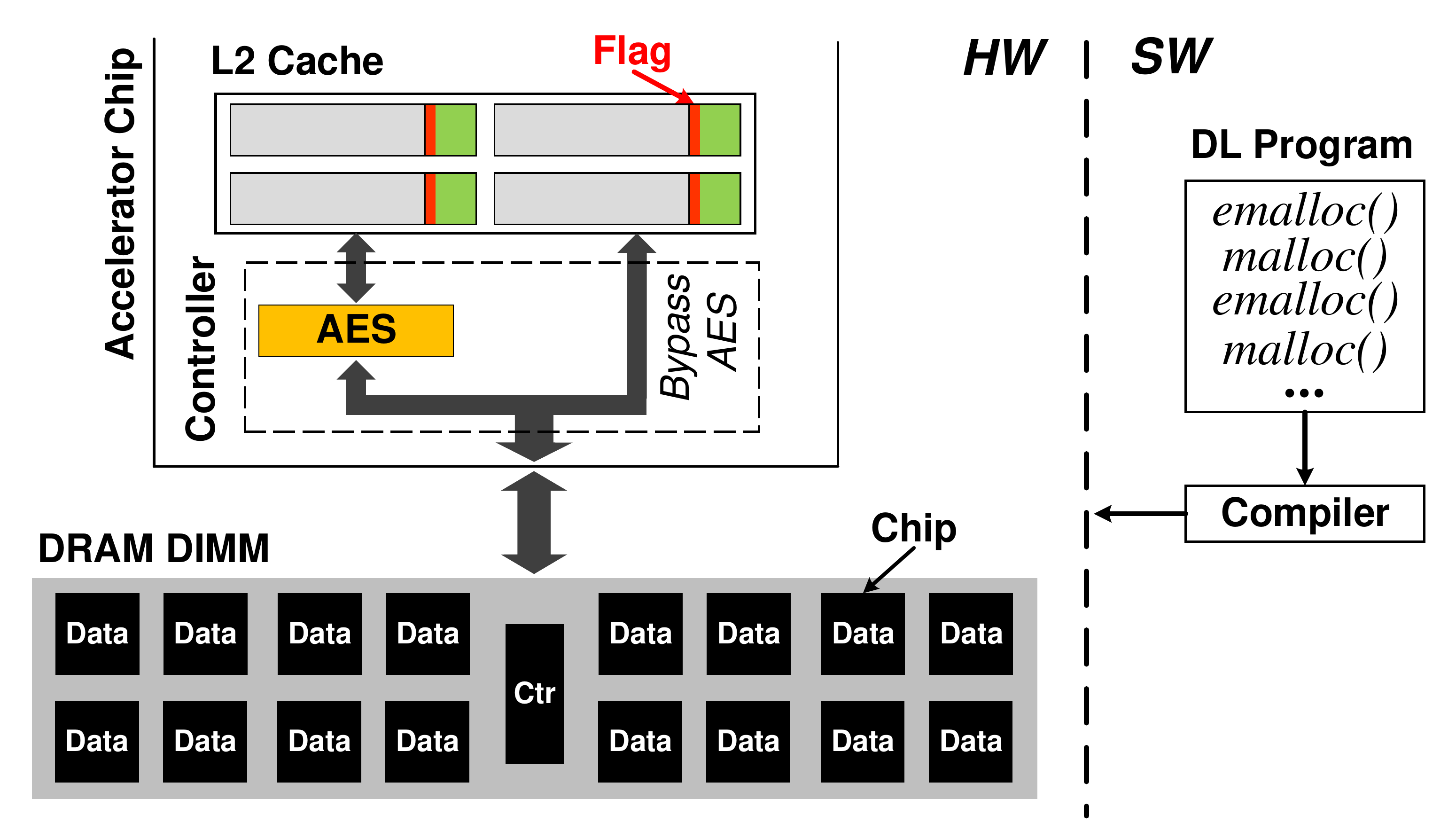}
    \vspace{-5px}
    \caption{\label{fig:3-7-SEAL-architecture} \textbf{A high-level overview of the SEAL architecture.} \textit{(This figure shows one memory controller and the other controllers are the same as this one.)}}
    \vspace{-20px}
\end{figure}

To support the proposed SE and ColoE schemes, SEAL is implemented via exploring and exploiting software and hardware co-design. The implementation and overall architecture of SEAL are shown in Figure~\ref{fig:3-7-SEAL-architecture}.

\isca{To support the SE scheme, in the software layer, we expose a new programming primitive,} \verb"emalloc()", \isca{to the high-level program in order to allow programmers to leverage the benefits of the smart encryption. The memory space allocated by} \verb"emalloc()" \isca{needs to be encrypted. The memory space allocated by existing} \verb"malloc()" \isca{in current programming languages does not need to be encrypted. In the hardware layer, the counter area is 64 bits while the counter used in the counter mode encryption only needs 56 bits, like the implementation in Intel's SGX~\cite{cryptoeprint2016204,gueron2016memory}. Thus 8 bits in the counter area are not used. We use one bit in the counter area of each memory line as a flag to indicate whether the memory line is allocated by} \verb"emalloc()" or \verb"malloc()". \isca{Hence, memory controllers can distinguish the memory lines allocated by} \verb"emalloc()" or \verb"malloc()" \isca{based on the flags. Memory lines allocated by} \verb"malloc()" \isca{bypass the AES engine.}

\isca{To support the ColoE scheme, referring to the design of error-correcting code (ECC) DRAM~\cite{chen1984error,dell1997white}, we design the DRAM DIMM to include an extra chip without changing the DRAM burst mechanism. As shown in Figure~\ref{fig:3-7-SEAL-architecture}, in a DRAM rank, there are 16 data chips and 1 counter chip (in the ECC DRAM, the chip is used for storing ECC bits). For a memory line, 128B data is stored in the 16 data chips (8B per chip) and 8B counter area is stored in the counter chip.}

\vspace{-3px}
\subsection{Security Analysis}
\label{sec:security-analysis}

For the security analysis, we first discuss the case where an adversary does not know what NN architecture is used in the target DL accelerator. In this case, even though some NN model data are obtained by the bus snooping attack, the adversary is difficult to distinguish which data are used for a particular layer. In our proposed SE scheme, some data are encrypted and hence it is more difficult for the adversary to recover the NN model. Therefore, we consider a strong attack model in which an adversary is able to figure out the NN architecture in the DL accelerator via side channel information~\cite{hua2018reverse,hu2019neural,yan2018cache},e.g., memory access patterns obtained from the memory bus, or device specifications~\cite{intel2017DLinference}. In this case, the adversary can distinguish the data from different layers and know the locations in the NN model where the encrypted and unencrypted data correspond to. Under the strong attack model, we below present the security analysis.	
The security of NN models involves two aspects including IP stealing and adversarial attacks, as presented in Section~\ref{background-adversarial-attacks}.

\subsubsection{Substitute Model Generation}
\label{sec:security-methodology}

\postmicro{In the security evaluation tests, we use three classical CNN models including VGG-16~\cite{simonyan2014very}, ResNet-18~\cite{he2016deep}, and ResNet-34~\cite{he2016deep} and train them on the widely used CIFAR-10 dataset~\cite{krizhevsky2009learning}. The NN model stored in the target DL accelerator is called \emph{victim model}, and the NN model that the adversary extracts from the accelerator by using bus-snooping attacks is called \emph{substitute model}.
Based on the fact that the adversary does not know the training dataset of the victim model, we isolate 90\% of training samples (45,000 images) in CIFAR10 as the training dataset of the victim model~\cite{Papernot2017PBA}. The remaining 10\% of training samples (5,000 images) are used by the adversary. Based on the 5,000 images, the adversary uses Jacobian-based dataset augmentation~\cite{Papernot2017PBA} to generate additional 40,000 images and then query them in the target accelerator to obtain their corresponding labels. The generated image-label pairs are used as the training dataset of the adversary's substitute models.
There are three kinds of substitute models that the adversary may obtain as follows.}

  \postmicro{\emph{$\bullet$ White-box model.} If a DL accelerator does not equip memory encryption, the adversary can know the entire victim model including all weight parameters and the NN architecture. Thus we consider an NN model that is the same as the victim model as the white-box substitute model.}

  \postmicro{\emph{$\bullet$ Black-box model.} If we encrypt all the victim model data and intermediate data, the adversary knows the NN architecture but does not know any weight parameters. However, the adversary can feed his/her own images into the target DL accelerator and obtain the output label. By using the image-label pairs, the adversary is able to retrain an NN model with the same architecture as the victim model. We consider the retrained NN model as the black-box substitute model.}

\begin{figure}[t]
	 \vspace{-8px} 
  \centering
    \includegraphics [width=0.42\textwidth]{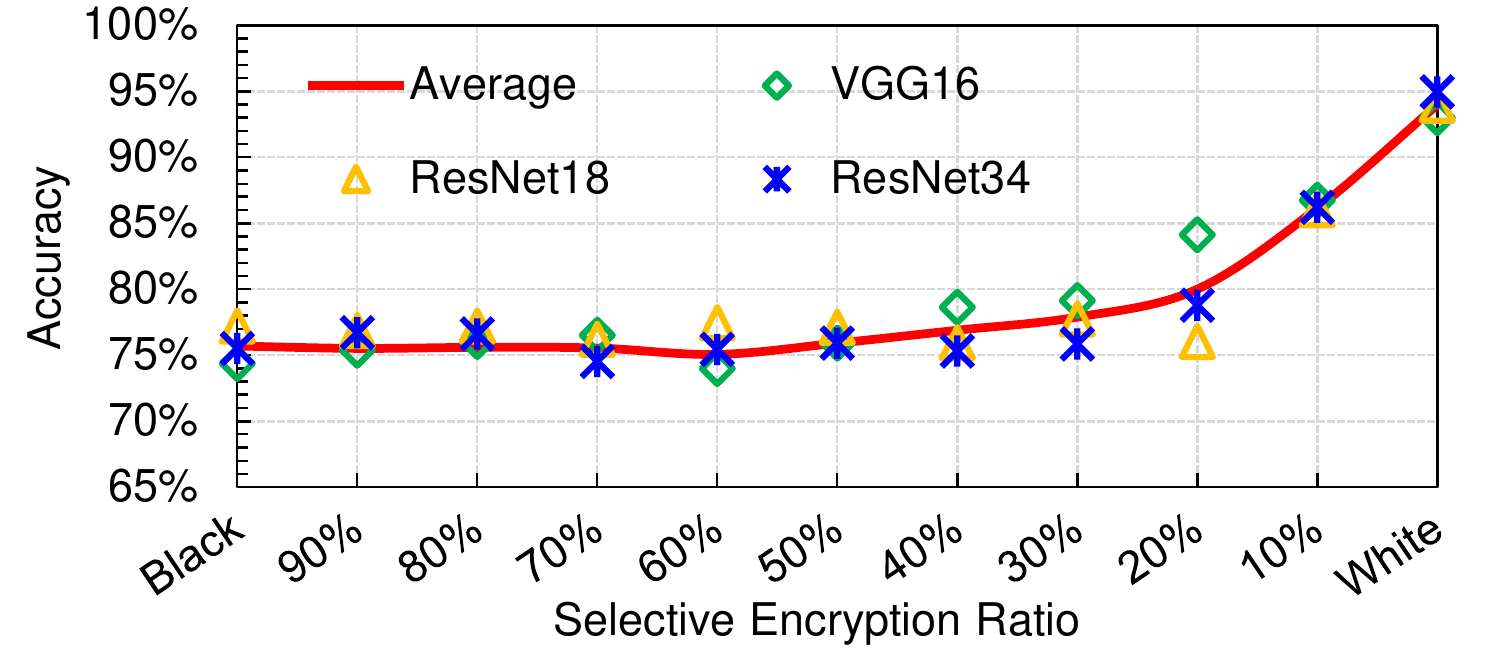}
    \vspace{-5px}
    \caption{\label{fig:design-IP-protection} \textbf{The inference accuracy of substitute models.}}
    \vspace{-19px}
\end{figure}

  \postmicro{\emph{$\bullet$ Smart encryption (SE) models.} In the SEAL, we selectively encrypt partial data that are critical and thus the adversary knows the NN architecture and partial weight parameters that are unencrypted. We perform full encryption on the first two CONV layers, the last one CONV layer, and the last FC layers of a CNN model to prevent the adversary from calculating the weight parameters via input and output layers, and perform the SE scheme on the remaining weight layers. However, by using inputs and outputs of the target DL accelerator, the adversary is able to supplement the unknown part of weight parameters via retraining the NN. Specifically, the adversary initializes an NN model with known weight parameters and fills random numbers following a standard normal distribution for unknown weight parameters~\cite{he2015delving}. The adversary then keeps the known weight parameters unchanged and fine-tunes unknown weight parameters by retraining the NN using inputs and outputs of the target DL accelerator. Note that the attacker can know the information that the sums of unknown weight rows must be larger than those of known weight rows and then leverage this information during fine-tuning. However, in our experiments, we observe the generated substitute models leveraging the information do not perform better, since limiting the sums of unknown weight rows may destroy efficient parameter fine-tuning.}

\subsubsection{Security on IP Stealing}
\label{sec:security-IP-protection}

\postmicro{
	One of the attack purposes is to steal the IP of NN models. The adversary that may be a business competitor aims to reduce the competitive advantages of model owners.
	The efficiency of the stolen attacks depends on the inference accuracy of the extracted substitute models. In the stolen attack tests, we first generate the three kinds of substitute models including white-box, black-box, and SE models that the adversary may obtain as mentioned above. For the SE models, we vary the encryption ratio from 90\% to 10\%. The encryption ratio is defined as the ratio of encrypted weight parameters to all weight parameters in each layer. The encrypted weights have the largest absolute weight values in each layer as presented in Section~\ref{sec:selective-encryption-scheme}. We evaluate the inference accuracy of these substitute models using test samples of the victim model.}

\begin{figure}[t]
    \vspace{-6px}
  \centering
    \includegraphics [width=0.45\textwidth]{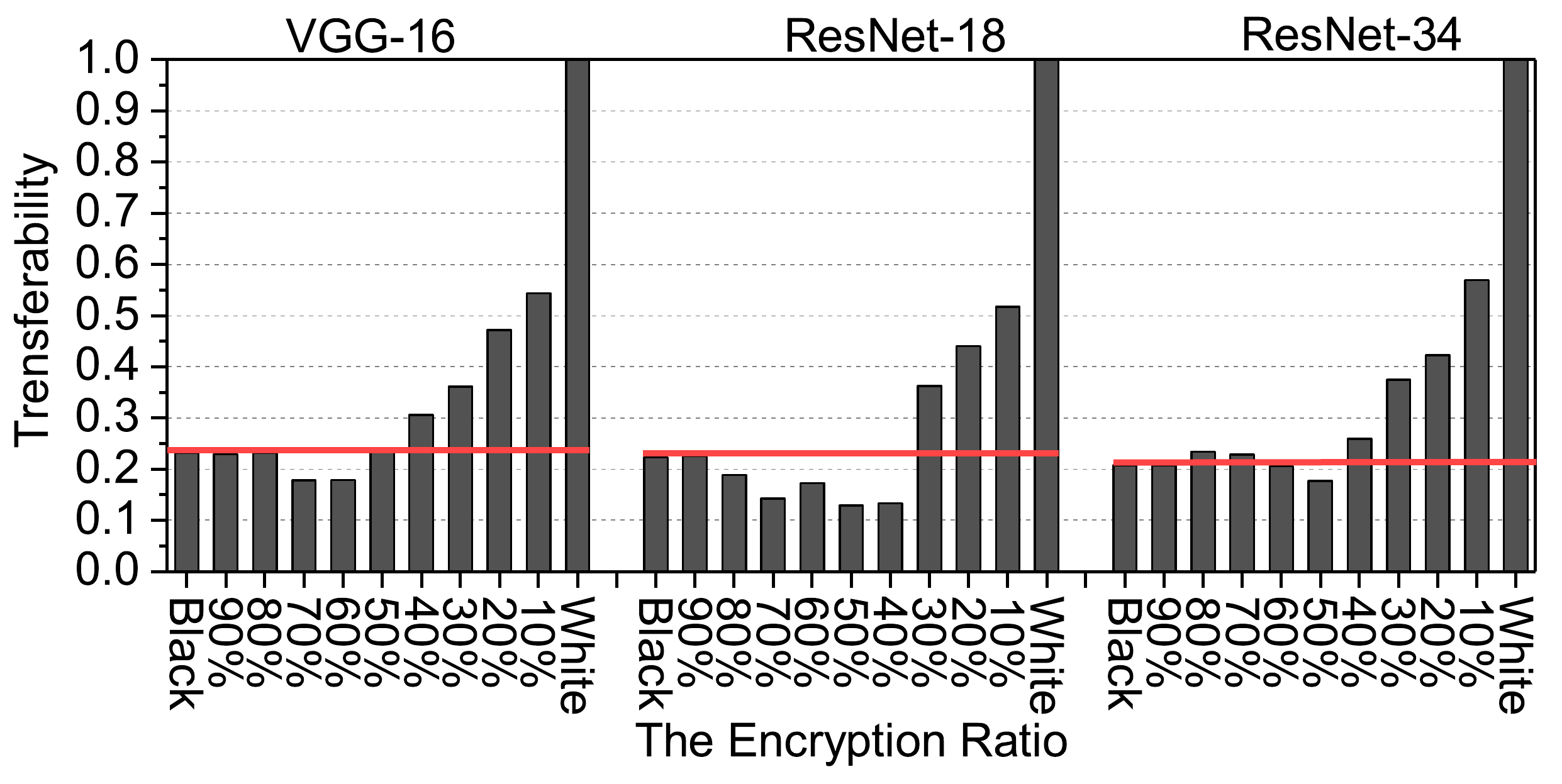}
    \vspace{-6px}
    \caption{\label{design-success-ratio} \textbf{The transferability of adversarial attacks for different substitute models.}}
    \vspace{-12px}
\end{figure}

\postmicro{Figure~\ref{fig:design-IP-protection} shows their inference accuracy. We observe that the while-box model has a very high accuracy, i.e., about 94\%, due to being the same as the victim model. The black-box model significantly reduces the accuracy from 94\% to 75\%. This is because the adversary does not know any weights and training samples in the victim model, and the black-box model can only be trained from a blank model by using the adversary's training dataset. For the SE models, when the encryption ratio is only 20\%, the accuracy significantly decreases by 14\% on average (from 94\% to 80\%), since the weight parameters with the largest absolutes are encrypted in SE models. When the encryption ratios $\geq 40\%$, the accuracy is almost the same as that of the black-box model. It means the SEAL with a $\geq 40\%$ encryption ratio achieves the same security level as the black-box model for IP protection.}

\subsubsection{Security on Adversarial Attacks}
\label{sec:security-adversatial-attacks}

\postmicro{If the purpose is to attack the victim model, the adversary can use the extracted NN models to generate adversarial examples and then use the adversarial examples to perform adversarial attacks.
In the adversarial attacks, the adversary aims to add the minimum perturbation on the input to mislead the victim model to produce a pre-assigned incorrect output~\cite{kurakin2016adversarial,alfeld2016data,szegedy2014intriguing}.
In the adversarial attack tests, we use the three kinds of substitute models including white-box, black-box, and SE models to respectively generate 1,000 adversarial examples via the I-FGSM method~\cite{kurakin2016adversarial}. Each batch of 1,000 adversarial examples have a 100\% attack success rate to attack their corresponding substitute models. We then use these adversarial examples to attack the victim model and evaluate the transferability of adversarial examples. The transferability is defined as the ratio of the adversarial examples that successfully attack the victim model to all adversarial examples, which is a widely used metric to evaluate the efficiency of substitute models for adversarial attacks~\cite{zhao2019compress,tramer2018ensemble,liu2017delving,goodfellow2018making}. Figure~\ref{design-success-ratio} shows the transferability of the adversarial examples generated by different substitute models.}

We observe that black-box models have much low transferability (about 20\%) for the three CNN models compared with while-box models, since the adversary with black-box models does not know any weight parameters and training samples of the victim model. For the SE models, when the encryption ratios $\geq$ 50\% for the three CNN models, the transferability is close to, and even smaller than those of black-box models. The reason is that the unencrypted weight parameters in the SE scheme are relatively un-important because they have the smallest absolute weights in each layer. If the adversary keeps the unencrypted weight parameters unchanged and fine-tunes the remaining weight parameters, the unchanged, unimportant weight parameters may disturb the retrained model, thus producing smaller attack success rates than the black-box model. When the encryption ratios $<$ 40\%, the transferability rapidly increases since some important weight parameters with large absolutes are exposed to the adversary.
Based on the above results, we set the encryption ratio of the SE scheme to 50\%, which obtains the maximum performance benefit when achieving the same security level as the black-box models.

%% file: evaluation.tex
\vspace{-4px}
\section{Performance Evaluation}
\label{sec:evaluation-methodology}
\vspace{-4px}
\subsection{Methodology}
\label{sec:evaluation-methodology-config}

We evaluate the performance of SEAL using the GPGPU-Sim v3.2.2~\cite{bakhoda2009analyzing}, a cycle-level simulator for contemporary GPUs. We model the microarchitecture for NVIDIA GeForce GTX480 GPU~\cite{Nvidia2012GTX480} with 15 streaming multiprocessors (SMs), one of the default GPUs in GPGPU-Sim. The details of our used GPU configuration are shown in Table~\ref{tab:gpu-configuration}. \postmicro{Although we perform our simulations on an Nvidia Fermi GPU, our solution focusing on the accelerator memory system is also applicable and generalizable to newer GPU architectures including Maxwell, Pascal, and Volta}, as well as other kinds of DL accelerators as presented in Section~\ref{sec:background-straightforward-solution}.
To implement SEAL, we add an AES encryption engine in every memory controller of the simulated GPU. We model a pipeline AES encryption engine with 128-bit block~\cite{mathew201053gbps,Awad2017OLA,opencores2012aes}, in which the overall AES encryption latency for a cache line is 20 cycles and \iscare{the bandwidth of an AES engine is 8GB/s. According to bandwidth values summarized in Tables~\ref{tab:backgorund-banwidth} and~\ref{tab:AES-engines}, we set a mean bandwidth value for AES and GPU.}

\begin{table}[t]
\caption{\textbf{Configurations of the simulated system.}}
\vspace{-12px}
\label{tab:gpu-configuration}
\scriptsize
\begin{center}
   \begin{tabular}{|l|l|}
    \hline
    \multicolumn{2}{|c|}{\textbf{GPU Core}}  \\
    \hline
     Number of SMs & 15 \\
    \hline
        Core clock & 700 MHz \\
    \hline
    Number of warps per SM & 48 \\
    \hline
    Register file size per SM & 128KB (32768 registers)\\
    \hline
    Register file cache size per SM & 16KB (4096 registers)  \\
    \hline
    Shared memory size per SM &  48KB \\
    \hline
    \multicolumn{2}{|c|}{\textbf{Cache and Memory}}  \\
    \hline
    Private L1 cache & 16KB, 4-way, 128B line, 1-cycle latency \\
    \hline
    Shared L2 cache & 768KB, 8-way, 128B line, 10-cycle latency \\
    \hline
    \multirow{2}{*}{Memory model} & GDDR5, 1848 MHz (3696 data rate), \\
    & 384-bit bus width, 6 channels, FR-FCFS \\
    \hline
   \multirow{2}{*}{GDDR5 timing (ns)} & $t_{CL}$ = 12, $t_{RP}$ = 12, $t_{RC}$ = 40,\\
   & $t_{RAS}$ = 28, $t_{RCD}$ = 12, $t_{RRD}$ = 6 \\
   \hline
\end{tabular}
\end{center}
\vspace{-20px}
\end{table}

\textbf{Benchmarks.} We use three classical CNN models including VGG-16~\cite{simonyan2014very}, ResNet-18~\cite{he2016deep}, and ResNet-34~\cite{he2016deep}. In order to run these CNN models on GPGPU-Sim, we install the PyTorch for GPGPU-Sim~\cite{pytorchforgpusim2018}, an open-source modified version of PyTorch that enables GPGPU-Sim to use the cuDNN library~\cite{cublas2007}.

\textbf{Comparisons.} We compare SEAL with the five schemes.
\begin{itemize}
  \vspace{-4px}
  \item \verb"Baseline": An insecure GPU without memory encryption as the baseline.
  \vspace{-3px}
  \item \verb"Direct": A straightforward solution using the direct encryption scheme as presented in Section~\ref{sec:background-straightforward-solution}.
  \vspace{-3px}
  \item \verb"Counter": A straightforward solution using the counter mode encryption scheme as presented in Section~\ref{sec:background-straightforward-solution}. For the counter mode encryption scheme, we add an on-chip counter cache whose size is 1/16 (equal to the counter/data size ratio, i.e., 8B/128B) of the L2 cache.
  \vspace{-3px}
  \item \verb"Direct+SE": The direct encryption scheme with our proposed criticality-aware smart encryption (SE) scheme. We compare the performance of \verb"Direct" and \verb"Direct+SE" to show the benefit of the SE scheme.
  \vspace{-3px}
  \item \verb"Counter+SE": The counter mode encryption scheme with our proposed SE scheme. We compare the performance of \verb"Counter" and \verb"Counter+SE" to show the benefit of the SE scheme. We also compare \verb"Counter+SE" with \verb"SEAL" to show the benefit of our proposed colocation mode encryption (ColoE) scheme.
  \vspace{-5px}
\end{itemize}


\vspace{-3px}
\subsection{Performance of Different Layers}
\vspace{-3px}

\subsubsection{IPC of Different-layer Computation}

We perform the SE scheme on CONV layers whose input and output feature maps are also the input and output of POOL layers.
Different encryption schemes have different impacts on the performance of CONV and POOL layers.
The default encryption ratio is 50\% for the SE scheme as presented in Section~\ref{sec:security-analysis}.
To investigate the impact of different encryption schemes on the performance of different layers, we evaluate four typical CONV layers in VGG, in which the number of input and output channels is 64, 128, 256, and 512, respectively. We also evaluate the five different POOL layers.

\begin{figure}[t]
    \vspace{-3px}
  \centering
    \includegraphics [width=0.41\textwidth]{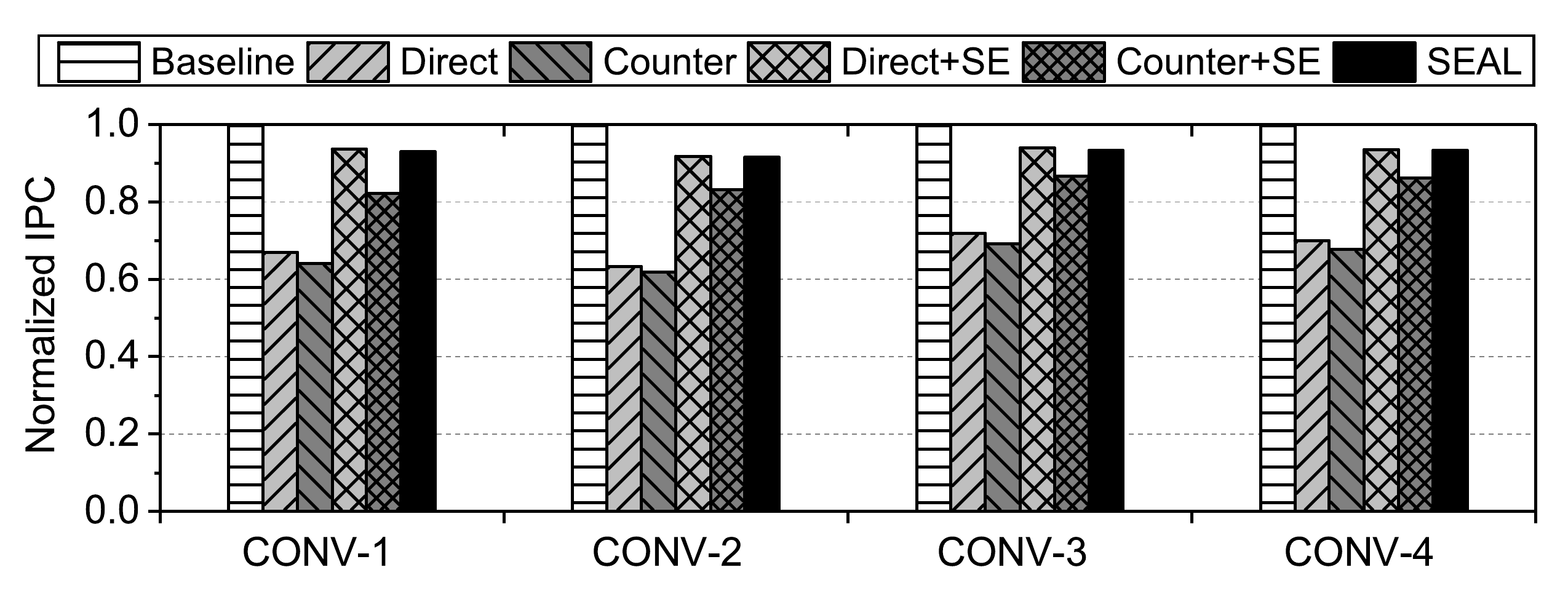}
    \vspace{-7px}
    \caption{\label{fig:evaluation-conv-ipc} \textbf{The IPC of different encryption schemes normalized to that of a baseline GPU for CONV layers.}}
    \vspace{-6px}
\end{figure}

Figure~\ref{fig:evaluation-conv-ipc} shows the relative IPCs of different encryption schemes when computing these CONV layers. We observe that the Direct scheme and the Counter scheme reduce the GPU IPC by up to $40\%$ compared with the baseline GPU without memory encryption. The reason is that memory encryption significantly reduces the data access bandwidth in GPUs as discussed in Section~\ref{sec:background-straightforward-solution}.
By comparing the performance between the Direct/Counter and the Direct+SE/Counter+SE schemes, our proposed SE scheme significantly improves the memory encryption performance on GPUs by reducing the amount of the encrypted data to improve the data access bandwidth without compromising security. The Direct+SE scheme has higher IPC performance than the Counter+SE scheme, since the counter mode encryption causes extra memory accesses from counters. However, the direct encryption has a lower security level than the counter mode encryption. SEAL leverages a ColoE scheme to achieve the same security level as counter mode encryption while delivering higher performance. Compared with the Counter+SE scheme, we observe that SEAL improves the IPC by up to 12\% by using the ColoE scheme.

\begin{figure}[t]
\vspace{-3px}
  \centering
    \includegraphics [width=0.41\textwidth]{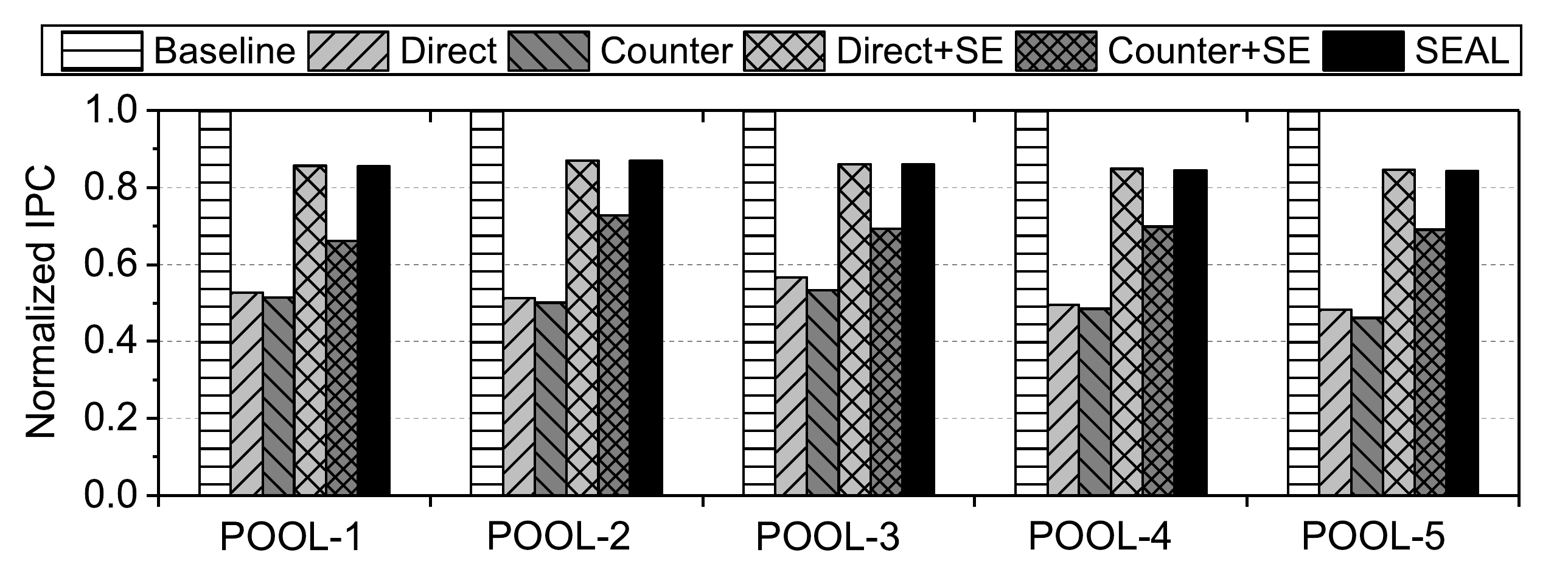}
    \vspace{-7px}
    \caption{\label{fig:evaluation-pool-ipc} \textbf{The IPC of different encryption schemes normalized to that of a baseline GPU for POOL layers.}}
     \vspace{-9px}
\end{figure}

Figure~\ref{fig:evaluation-pool-ipc} shows the relative IPCs of different encryption schemes when computing POOL layers. We observe the Direct and Counter schemes reduce the IPC by up to $50\%$, and perform worse in comparison to computing CONV layers since the computation of POOL layers is more bandwidth-bounded than that of CONV layers. Due to the same reason, the Direct+SE, Counter+SE, and SEAL perform worse when compared to computing CONV layers. Nevertheless, for the entire neural network, the amount of computation overhead in CONV layers is much larger than that in POOL layers.

\begin{figure}[t]
  \centering
    \includegraphics [width=0.41\textwidth]{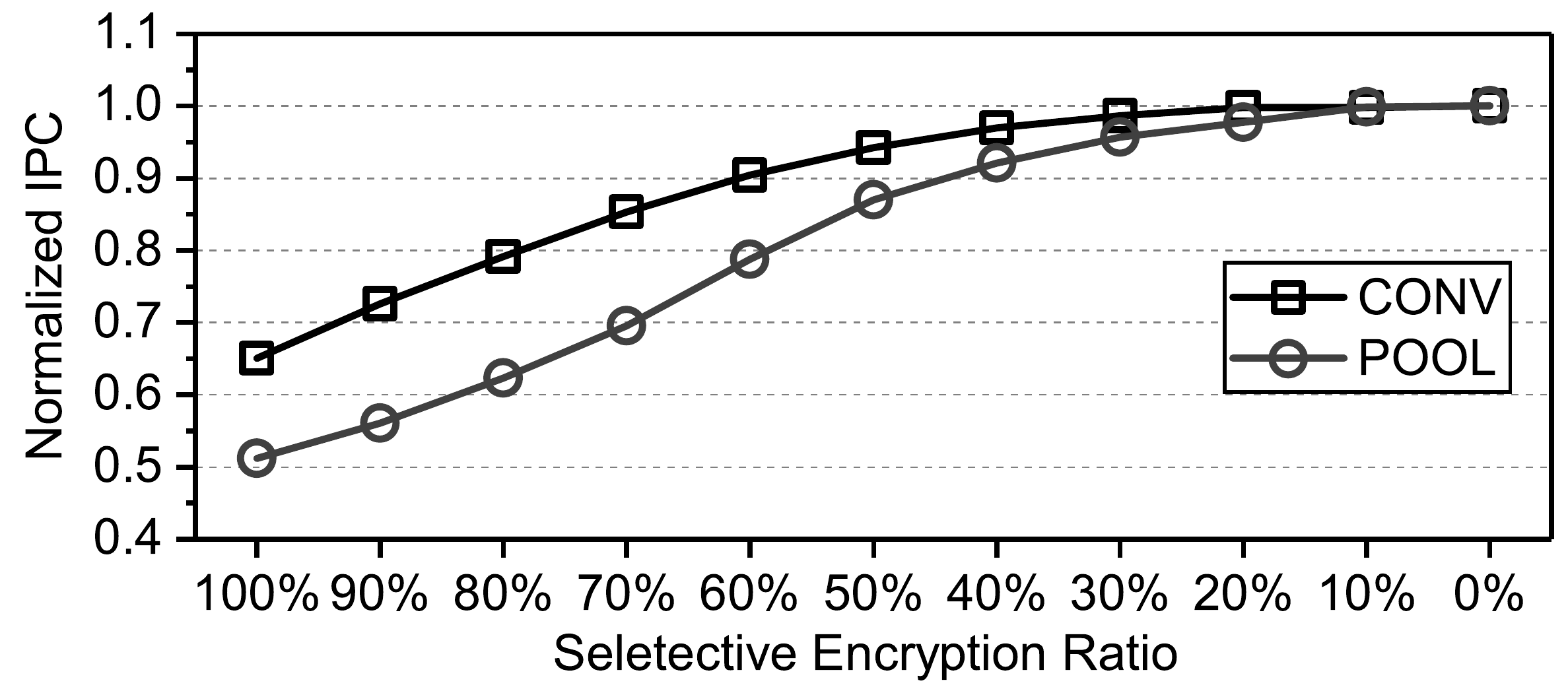}
    \vspace{-6px}
    \caption{\label{fig:evaluation-different-se-ipc} \textbf{The IPC of SEAL with different encryption ratios normalized to that of a baseline GPU.}}
    \vspace{-13px}
\end{figure}

\subsubsection{Performance under Different Encr. Ratios}

We investigate the impact of different encryption ratios on the performance of SEAL when computing a CONV/POOL layer. We vary the encryption ratio from 100\% to 0\% with the 10\% interval. A 100\% encryption ratio means a full-encryption GPU. When the encryption ratio is 0\%, the performance is the same as that of a baseline GPU without memory encryption. The experimental results are shown in Figure~\ref{fig:evaluation-different-se-ipc}. When slightly reducing the encryption ratio by $20\%-30\%$ from a 100\% encryption ratio, the IPC is significantly improved. The reason is that allowing partial data to bypass the AES encryption engine not only improves the data access bandwidth of the GPU but also reduces the competition for the use of the encryption engine. When the encryption ratio is reduced to 50\%, the IPC is improved from 65\% to 95\% and from 54\% to 87\% for computing the CONV and POOL layers respectively, compared with a 100\% encryption ratio.

\vspace{-2px}
\subsection{Overall Performance}
\vspace{-4px}
\subsubsection{IPC}
\begin{figure}[h]
	\vspace{-6px}
  \centering
    \includegraphics [width=0.41\textwidth]{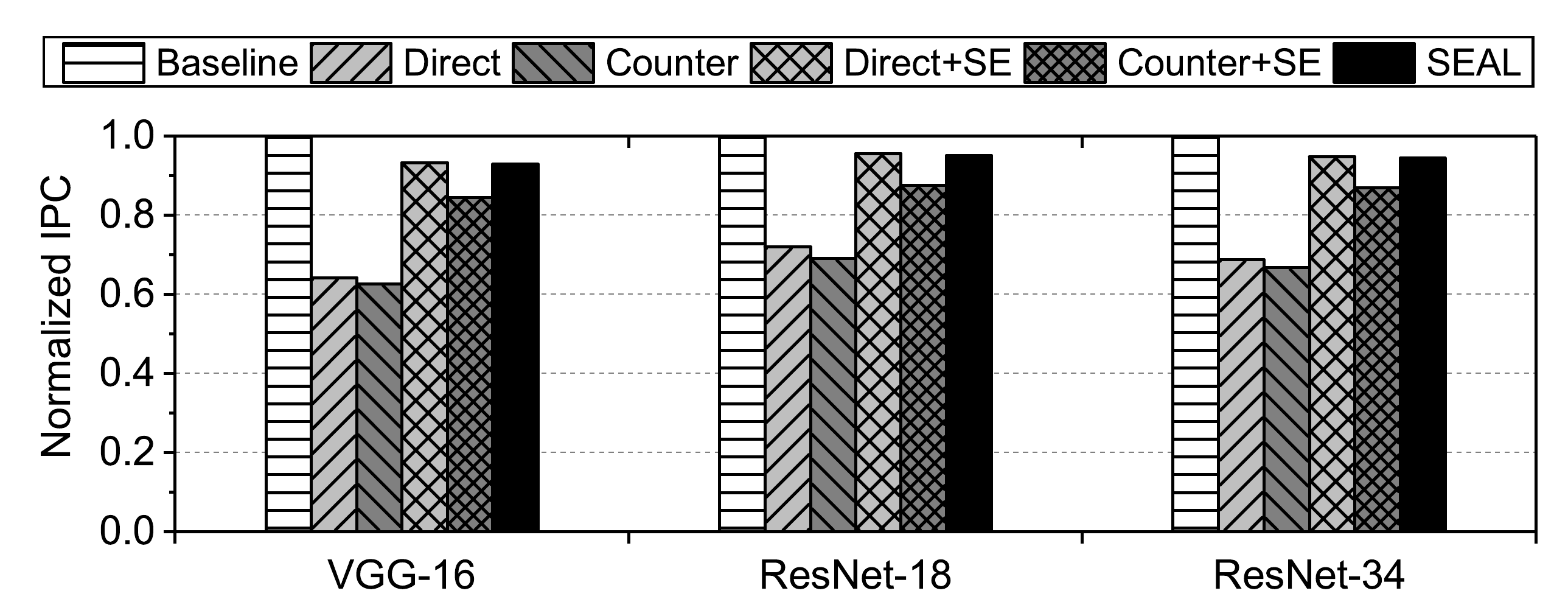}
    \vspace{-6px}
    \caption{\label{fig:evaluation-overall-ipc} \textbf{The IPC normalized to that of a baseline GPU.}}
    \vspace{-13px}
\end{figure}

We evaluate the IPC of the GPU with different encryption schemes when executing the NN inference using VGG-16, ResNet-18, and ResNet-34, as shown in Figure~\ref{fig:evaluation-overall-ipc}.
Traditional memory encryption solutions including the Direct and Counter schemes reduce the GPU IPC for executing NN inference by $30\%-38\%$, compared with a baseline GPU. Moreover, the Direct and Counter schemes deliver higher performance in ResNets than those in VGG. The reason is that the amounts of computation and data accesses to memory in VGG are much larger than those in ResNets~\cite{he2016deep} and thus VGG requires higher data access bandwidth. Memory encryption limits the data access bandwidth and hence has more significant impact on the performance of VGG. By using our proposed SE scheme to allow some data to bypass the AES encryption engine, the Direct+SE and Counter+SE schemes improve the IPC by about $31\%$ and $20\%$ respectively, compared with the Direct and Counter schemes. By using our proposed ColoE scheme to eliminate memory accesses of counters, SEAL further improves the IPC by about 7\% compared with the Counter+SE scheme. As a result, compared with the traditional memory encryption solutions, i.e., the Direct and Counter schemes, SEAL achieves $1.4\times-1.6 \times$ IPC improvement. Moreover, SEAL achieves the $93\%-95\%$ IPC of a baseline GPU without memory encryption, i.e., compromising only $5\%-7\%$ performance for security \postmicro{improvement}.

\begin{figure}[t]
  \centering
    \includegraphics [width=0.38\textwidth]{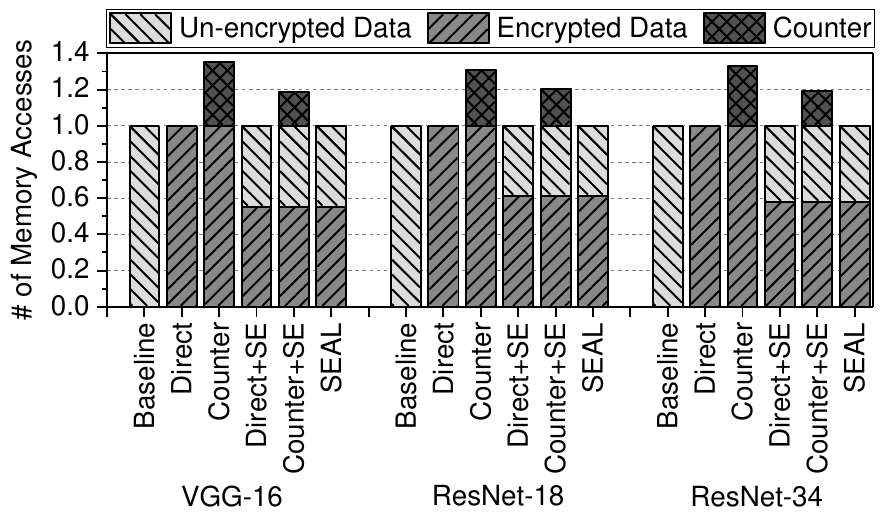}
    \vspace{-9px}
    \caption{\label{fig:evaluation-overall-memory-access} \textbf{The number of memory accesses normalized to that of a baseline GPU.}}
    \vspace{-11px}
\end{figure}

\subsubsection{The Number of Memory Accesses}
We evaluate the number of different kinds of memory accesses when using different encryption schemes, as shown in Figure~\ref{fig:evaluation-overall-memory-access}. For the baseline GPU, all memory accesses including reads and writes come from unencrypted data. For the Direct scheme, all memory accesses are from encrypted data and thus need to pass through the low-bandwidth AES encryption engine. Therefore, the Direct scheme significantly reduces the GPU performance compared with the Baseline as shown in Figure~\ref{fig:evaluation-overall-ipc}. For the Counter scheme, all memory accesses from data are also from encrypted data. Moreover, the Counter scheme incurs $31\%-35\%$ more memory accesses from counters and thus has lower performance than the Direct scheme. Nevertheless, in the Direct and Counter schemes, the main performance bottleneck is the AES encryption engine rather than the DRAM. Hence, extra memory accesses from counters in Counter scheme do not incur much performance decrease, compared with the Direct scheme.

By using the SE scheme, the number of memory accesses from encrypted data is reduced by $39\%-45\%$. Therefore, the IPCs of Direct+SE and Counter+SE schemes are significantly improved compared with the Direct and Counter schemes, as shown in Figure~\ref{fig:evaluation-overall-ipc}. Moreover, the Counter+SE scheme incurs about $20\%$ more memory accesses from counters and thus has lower performance than the Direct+SE scheme. Nevertheless, in the Direct+SE and Counter+SE schemes, the AES encryption engine may not be the main performance bottleneck due to using the SE scheme. Therefore, extra $20\%$ memory accesses from counters in Counter scheme incur much performance decrease, compared with the Direct scheme. SEAL leverages the ColoE scheme to achieve the same security level as counter mode encryption while eliminating the extra memory accesses from counters. Therefore, compared with the Direct+SE scheme, SEAL achieves higher security level and the approximate performance. Compared with the Counter+SE scheme, SEAL achieves higher performance and the same security level.

\vspace{-2px}
\subsubsection{Inference Latency}
\begin{figure}[t]
  \centering
    \includegraphics [width=0.41\textwidth]{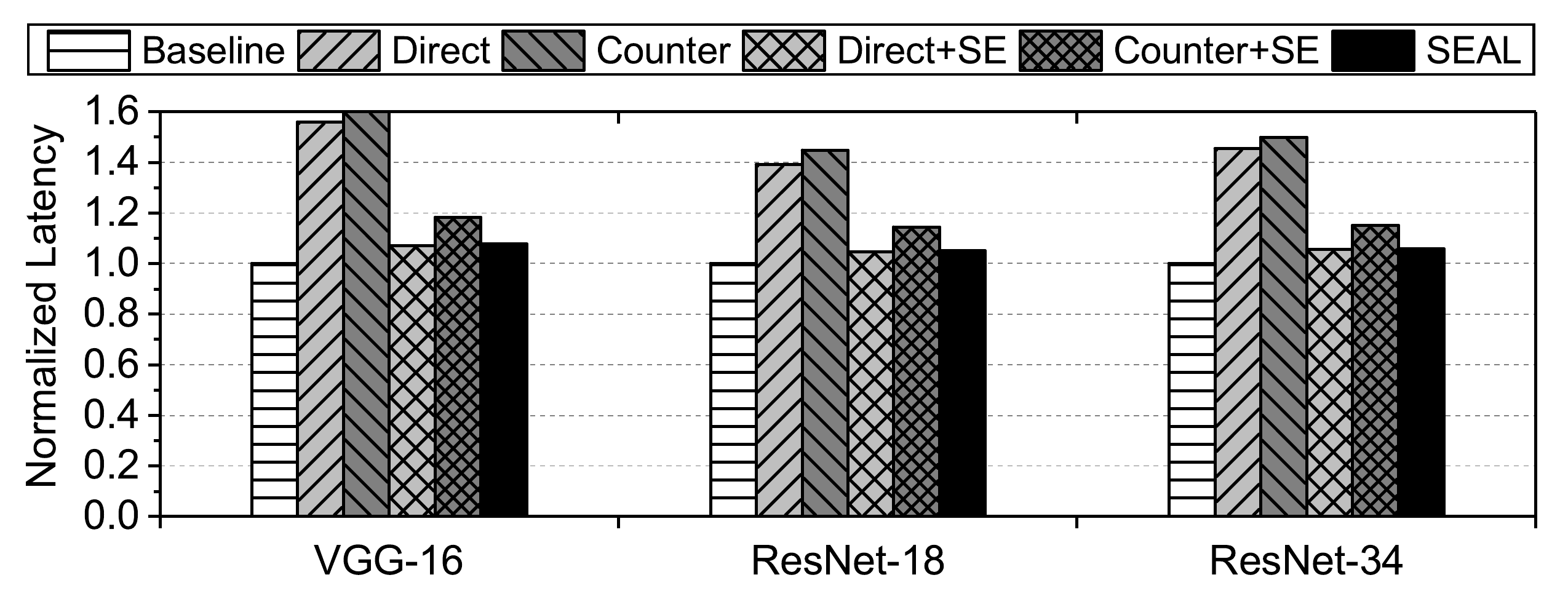}
    \vspace{-8px}
    \caption{\label{figevaluation-overall-latency} \textbf{The inference latency normalized to that of a baseline GPU.}}
    \vspace{-16px}
\end{figure}

We investigate the impact of different encryption schemes on the inference latency, as shown in Figure~\ref{figevaluation-overall-latency}. Traditional memory encryption solutions including the Direct and Counter schemes increase the inference latency by $39\%-60\%$, compared to a baseline GPU. By using the proposed SE scheme, the Direct+SE and Counter+SE schemes reduce the extra inference latency to $5\%-18\%$. By using both the SE and ColoE schemes, SEAL incurs only $5\%-7\%$ higher inference latency than the baseline GPU.

%% file: relatedwork.tex
\vspace{-4px}
\section{Related Work}

\textbf{Model Extraction Attacks.} \emph{1) Algorithm layer.} There exist algorithm-layer approaches to extract the NN model related information by exploiting the inputs and outputs of the DL inference. Tram{\`e}r el al.~\cite{tramer2016stealing} assume that the confidence scores of the classification labels produced by DL systems and the NN model architecture are known and demonstrate that some model parameters can be speculated. Oh et al.~\cite{oh2018towards} assume the model architecture is unknown and propose a metamodel approach to extract the information of the NN model architecture. Wang et al.~\cite{wang2018stealing} propose an approach to extract the hyperparameters of the NN model. The hyperparameters are usually used to balance between the regularization terms in the objective function and the loss function. \emph{2) System and architecture layers.} Existing works exploit the information of the operating system and architecture layers to speculate the NN model related information. Naghibijouybari et al.~\cite{naghibijouybari2018rendered} exploit the side channel information in the operating system, such as memory allocation APIs, GPU performance counters, and timing measurement, to speculate the NN model related information, e.g., the number of neurons. Hua et al.~\cite{hua2018reverse} exploit the side channel information in the DL accelerator architecture, e.g., the memory access pattern, to speculate the NN architecture related information.

The model extraction attacks mentioned above can obtain only a small part of the NN model related information. Compared with these model extraction attacks, the bus snooping attacks for DL accelerators that our paper focuses on are much more dangerous. This is because an adversary can obtain all data of the entire NN model including weight parameters in each layer by the bus snooping attacks. Our paper proposes a secure and efficient solution, called SEAL, to defend against the bus snooping attacks for DL accelerators.

\textbf{Memory Encryption.} \postmicro{Obviously, software memory encryption, such as Graviton~\cite{volos2018graviton}, cannot adequately defend against physical access based attacks~\cite{Yan2006ICP}, since the programs of encryption software themselves can be stored in the memory.}
Hardware memory encryption has been widely used in secure CPU systems~\cite{Yan2006ICP,Young2015DWE,awad2016silent,lipmaa2000ctr,henson2014memory,saileshwar2018morphable} to defend against physical access based attacks by adding the hardware encryption engine on the CPU chip. Memory encryption does not cause significant performance degradation in CPU systems, since the DDR memory bus for CPUs has a similar bandwidth to the encryption engine. However, memory encryption significantly decreases the performance of DL accelerators, e.g., GPUs, due to the big bandwidth gap between the GDDR memory bus and encryption engine.
Our proposed SEAL is able to efficiently address this problem via criticality-aware smart encryption and colocation mode encryption.